\def\spose#1{\hbox to 0pt{#1\hss}}
\def\approxlt{\mathrel{\spose{\lower 3pt\hbox{$\sim$}}
        \raise 2.0pt\hbox{$<$}}}
\def\approxgt{\mathrel{\spose{\lower 3pt\hbox{$\sim$}}
        \raise 2.0pt\hbox{$>$}}}
\def\approxpropto{\mathrel{\spose{\lower 3pt\hbox{$\sim$}}
        \raise 2.0pt\hbox{$\propto$}}}
\mathchardef\twiddle="2218

\def\multleft#1{\hbox to size{\vbox {\halign {\lft{##}\cr #1}}\hfill}\par}
\def\multright#1{\hbox to size{\vbox {\halign {\rt{##}\cr #1}}\hfill}\par}
\def\Mdot{\hbox{$\dot M$}}

\def\<{\thinspace}

\def\erg{{\rm\thinspace erg}}

\def\K{{\rm\thinspace K}}
\def\keV{{\rm\thinspace keV}}

\def\km{{\rm\thinspace km}}
\def\kpc{{\rm\thinspace kpc}}

\def\Mpc{{\rm\thinspace Mpc}}
\def\Msun{\hbox{$\rm\thinspace M_{\odot}$}}

\def\s{{\rm\thinspace s}}
\def\yr{{\rm\thinspace yr}}

\def\ergps{\hbox{$\erg\s^{-1}\,$}}

\def\kmps{\hbox{$\km\s^{-1}\,$}}

\def\Msunpyr{\hbox{$\Msun\yr^{-1}\,$}}

\def\kmpspMpc{\hbox{$\kmps\Mpc^{-1}$}}

\long\def\symbolfootnote[#1]#2{\begingroup%
     \def\thefootnote{\fnsymbol{footnote}}\footnote[#1]{#2}\endgroup} 



\newcommand\beq{\begin{equation}}
\newcommand\eeq{\end{equation}}
\newcommand\beqa{\begin{eqnarray}}
\newcommand\eeqa{\end{eqnarray}}

\documentstyle[emulateapj,psfig,onecolfloat]{article}

\begin{document}

\twocolumn[

\title{Black hole growth and activity in a $\Lambda$CDM Universe}
\author{Tiziana Di Matteo\altaffilmark{1}, Rupert A.C. Croft\altaffilmark{2},
Volker Springel\altaffilmark{1}, Lars Hernquist\altaffilmark{3} }  

\altaffiltext{1}{Max-Planck-Institut f{\" u}r Astrophysik, Karl-Schwarzschild-Str.~1, 85740 Garching bei M{\" u}nchen, Germany}

\altaffiltext{2}{Carnegie-Mellon University, Dept. of Physics, 5000 Forbes Ave., Pittsburgh, PA 15213}  

\altaffiltext{3}{Harvard-Smithsonian Center for Astrophysics, 60
 Garden St., Cambridge, MA 02138} 

\begin{abstract}
The observed properties of supermassive black holes suggest a
fundamental link between their assembly and the formation of their
host spheroids. We model the growth and activity of black holes in
galaxies using $\Lambda$CDM cosmological hydrodynamic simulations
by following the evolution of the baryonic mass component in galaxy
potential wells. We find that the observed steep relation between
black hole mass and spheroid velocity dispersion, $M_{\rm BH}
\propto \sigma^4$, is reproduced if the gas mass in bulges is linearly
proportional to the black hole mass. To a good approximation, this
is equivalent to assuming the conversion of a fixed fraction of gas mass 
into black  hole mass. In this model, star formation 
and supernova feedback in the gas are
sufficient for regulating and limiting the growth of the central black
hole and of its gas supply. Black hole growth saturates because of the
competition with star-formation and in particular feedback, both of
which determine the gas fraction available for accretion. Unless other
processes also operate, we predict that the $M_{\rm BH}-\sigma$ relation
is not set in primordial structures but is fully established at low
redshifts, $z \approxlt 2$, and is shallower at earlier times.  Once
this relation is established we find that that central black hole
masses are related to their dark matter halos simply via $M_{\rm BH}
\approxpropto M_{\rm DM}^{4/3}$. We assume that galaxies undergo a
quasar phase with a typical lifetime, $t_{Q} \sim 2\times 10^7$ yr,
the only free parameter of the model, and show that star-formation
regulated depletion of gas in spheroids is sufficient to explain, for
the most part, the decrease of the quasar population at redshift $z
<3$ in the optical blue band. However, with the simplest assumption of
a redshift independent quasar lifetime, the model slightly
overpredicts optical quasar numbers at high redshifts although
it yields the observed evolution of number density of X-ray faint
quasars over the whole redshift range $1 < z< 6$.  Finally, we find
that the majority of black hole mass is assembled in galaxies by $z
\sim 3$ and that the black hole accretion rate density peaks in rough
correspondence to the star formation rate density at $z \sim 4-5$.
\end{abstract}

\keywords{accretion --- black hole physics ---
galaxy: evolution --- methods: numerical} ]

\altaffiltext{1}{Max-Planck-Institute f{\" u}r Astrophysik, Karl-Schwarzschild-Str. 1, 85740 Garching bei M{\" u}nchen, Germany}

\altaffiltext{2}{Carnegie-Mellon University, Dept. of Physics, 5000 Forbes Ave., Pittsburgh, PA 15213}  

\altaffiltext{3}{Harvard-Smithsonian Center for Astrophysics, 60
 Garden St., Cambridge, MA 02138} 

\section{Introduction}
Recent dynamical studies indicate that supermassive black holes reside
at the centers of most nearby galaxies (e.g., Kormendy \& Richstone
1995; Richstone et al. 1998; Kormendy \& Gebhardt 2001). The evidence
indicates that the mass of the central black hole is correlated with
the bulge luminosity (e.g., Magorrian et al. 1998) and even more
tightly with the velocity dispersion of its host bulge, where it is
found that $M_{\rm BH} \propto \sigma^{~4}$ (Tremaine et al. 2002;
Ferrarese \& Merritt 2000; Merritt \& Ferrarese 2001; Gebhardt et
al. 2000). The tight relation between the mass of the black holes and
the gravitational potential well that hosts them strengthens the
theoretical arguments that there is a fundamental link between the
assembly of black holes and the formation of spheroids in galaxy
halos. In addition, the locally inferred black hole mass density
appears to be broadly consistent with the density accreted during the
quasar phase (e.g; Soltan 1982; Fabian \& Iwasawa 1999; Yu \& Tremaine
2002) further supporting the idea that the nuclear activity, the
growth of the black holes and spheroid formation are all closely
linked.

It is not yet determined what fundamental physical process is
responsible for establishing the observed $M_{\rm BH} - \sigma$
relation or what drives the strong evolution of quasars at low
redshifts. Suggestions for the $M_{\rm BH} - \sigma$ dependence have
appealed to strong feedback due to quasar outflows on the
protogalactic gas reservoir (e.g., Silk \& Rees 1998); to capture of
stars following protogalactic collapse (Adams, Graff \& Richstone
2001) and to star formation regulated black hole growth (Burkert \&
Silk 2001). Models within the context of cold dark matter (CDM)
cosmologies, in which the black hole growth is linked to the growth of
galactic halos and activity is triggered during major mergers often
make use of the observed $M_{\rm BH} - \sigma$ relation (e.g., Wyithe
\& Loeb 2002; Volonteri, Haardt \& Madau 2002 and references therein),
or derive it dependent on some additional model assumptions (e.g.,
Kauffmann \& Haehnelt 2000; Haehnelt \& Kauffmann 2001).

In the commonly adopted merging scenario for the formation of
spheroids, it is well established that black hole masses resulting
from the coalescence of progenitor black holes fall short of the
measured values (e.g., Ciotti \& van Albada 2001). A large fraction of
the black hole mass must therefore have been accreted, presumably as
gas. At the same time, the gas present in the central region of
spheroids must provide the gas supply that will eventually be accreted
onto the black hole. In the simplest scenario the growth of
the central black hole and the amount of fuel available (hence the
black hole activity) should therefore be tied to the evolution of the
gas component in galaxies.

In this paper we investigate the growth and activity of black holes in
bulges and attempt to look for correlations with the large scale
properties of galaxies that match those of the observed black holes at
their centers. We use $\Lambda$CDM cosmological hydrodynamic
simulations to follow the evolution of the baryonic mass in the
potential wells of galaxies. The simulations (Springel \& Hernquist
2003a,b) include a new prescription for star formation and feedback
processes within the interstellar medium which has been shown to
produce a numerically converged prediction for the cosmic star
formation history.  Our aim is to trace from early times the gas
fraction available for accretion and hence for the growth and fueling
of black holes in competition with star formation and spheroid
growth. This should allow us to determine directly to what extent both the
$M_{\rm BH} - \sigma$ relation for black holes, and the
decline of quasars at low redshifts, are established through
self-regulated star formation processes in the popular cold dark
matter cosmogonies.

In \S 2 we briefly describe the simulations and our analysis.  In \S 3
and \S4 we present our results for the $M_{\rm BH} - \sigma$ relation
and our predictions for the luminosity function and space density
evolution of quasars. We discuss further implications of our results
in \S 5. In this paper we will limit ourselves to redshifts $z <7$
where the quasar luminosity function has been measured. In future work
we plan to extend our analysis to make predictions for the QSO
population up to very high redshifts.

\section{Simulations and analysis}
Throughout, we shall use a set of cosmological simulations for a
$\Lambda$CDM model, with $\Omega_{\Lambda}=0.7$, $\Omega_{\rm m}=0.3$,
baryon density $\Omega_{\rm b}=0.04$, a Hubble constant $H_{0} = 100 h 
\kmpspMpc $ (with $h=0.7$) and a scale-invariant primordial power spectrum
with index $n=1$, normalized to the abundance of rich galaxy clusters
at the present day ($\sigma_{8} =0.9$).  We here briefly summarize the
main features of our simulation methodology and refer to Springel \&
Hernquist (2003a,b) for a more detailed description.

Besides self-gravity of baryons and collisionless dark matter, the
simulations follow hydrodynamical shocks, and include radiative
heating and cooling processes of a primordial mix of helium and
hydrogen, subject to a spatially uniform, time-dependent UV background
(see, e.g., Katz et al. 1996, Dav\'e et al. 1999).  The dark matter
and gas are both represented computationally by particles.  In the
case of the gas, we use a smoothed particle hydrodynamics (SPH)
treatment (e.g., Springel et al. 2001) in its entropy formulation
(Springel \& Hernquist 2002).

In our numerical model, we assume that collapsed gas at very high
overdensity becomes available for star formation, which in turn
creates a complex multi-phase interstellar medium (ISM). We use a
sub-resolution model to describe the structure and dynamics of the ISM
on unresolved scales. In this model, star formation is assumed to
occur in cold clouds that form by thermal instability out of a hot
ambient medium, which is heated by supernova explosions. These
supernovae also evaporate clouds, thereby establishing a tight
self-regulation cycle for star formation in the ISM.

Motivated by observational evidence for the ubiquitous presence of
strong galactic outflows in actively star-forming galaxies (e.g.;
Martin et al. 1999; Heckman et al. 2000) and by the required
enrichment of the the interstellar medium (together with the
relatively low efficiency of the global star formation) we have also
included a phenomenological treatment of galactic winds. In the
fiducial parameterization of this process, each star forming region is
assumed to drive a wind with a mass outflow rate twice the star
formation rate. We then assume that the kinetic energy of the galactic
winds is comparable to the total available energy released by the
supernovae associated with star formation. This parametrization leads
to an initial wind speed in the simulations equal to $484\,{\rm
km\,s^{-1}}$. These wind parameters are chosen to be representative of
typical properties of the outflows associated with star forming disks
(e.g. Martin, 1999). We note that whether or not the wind will escape
from a galaxy (as it interacts and entrains infalling gas, shocks the
ISM etc.) depends primarily on the depth of the galaxy potential
well. For our choice of wind parameters we expect that only those
halos with virial temperatures below $10^6 \K$ have central escape
velocities lower than the wind speed and will then lose some baryons
in the outflows (c.f. \S 3.2).

Note that our simulations did not include any prescription for black
hole growth when they were run. The gas in our simulated galaxies is
therefore only affected by star formation and winds.

In order to resolve the full history of cosmic star formation,
Springel \& Hernquist (2003a,b) simulated a series of cosmological
volumes with sizes ranging from 1$h^{-1}$ Mpc to 100$h^{-1}$ Mpc.  For
each box size, a {\em series} of simulations was computed where the
mass resolution was increased systematically in steps of $1.5^3$, and
the spatial resolution in steps of 1.5.

In this work, we show results mostly from their `D-Series' which 
employed a periodic box of comoving side length 33.75 $h^{-1}$ Mpc. 
Within this
series, we use the `D3', `D4' and `D5' simulation runs with
resolutions of $2 \times 144^3$, $2\times 216^3$ and $2 \times 324^3$
particles, respectively, corresponding to mass resolutions in the gas
of $1.43 \times 10^{8} h^{-1}\Msun$, $4.24\times 10^7 h^{-1} \Msun$
and $1.26 \times 10^7 h^{-1} \Msun$. The spatial resolution of these
simulations can be characterized by their gravitational softening
lengths, which are equal to 9.38, 6.25, and 4.17 $h^{-1}
\kpc$ (in comoving units), respectively. 
The simulations of the D-Series have only been evolved to a minimum
redshift of $z=1$, because at this epoch, their fundamental mode
starts to become non-linear. Among the simulation program studied by
Springel \& Hernquist (2003b), the D-Series constitutes the best
compromise between redshift coverage and resolution appropriate for
the work here.  The analysis of three simulations of different
resolution but equal volume allows us to cleanly assess numerical
convergence for all the physical quantities we measure, which is a
significant advantage of our methodology.  For example, Springel \&
Hernquist (2003b) show that the star formation has fully converged
below $z \simeq 6$ in the D5 run, while this is only the case for
$z\simeq 5$ in D4, and for $z\simeq 4$ in D3.  In general, we
therefore expect that progressively higher resolution will also be
required to achieve convergence for measurements of other physical
quantities related to dissipative processes in the gas. We hence
analyze simulation outputs for a set of 10 redshifts, given by $z=7$,
$6$, $5$, $4$, $3.5$, $3$, $2.5$, $2$, $1.5$, and $1$.

Note that in this work we attempt to link the large-scale properties
of galaxies to those of observed black holes; for this goal it is not
important to numerically resolve the properties of the accretion flows
close to black holes.

\begin{figure*}[t]
\centerline{
\vbox{
\hbox{
\psfig{file=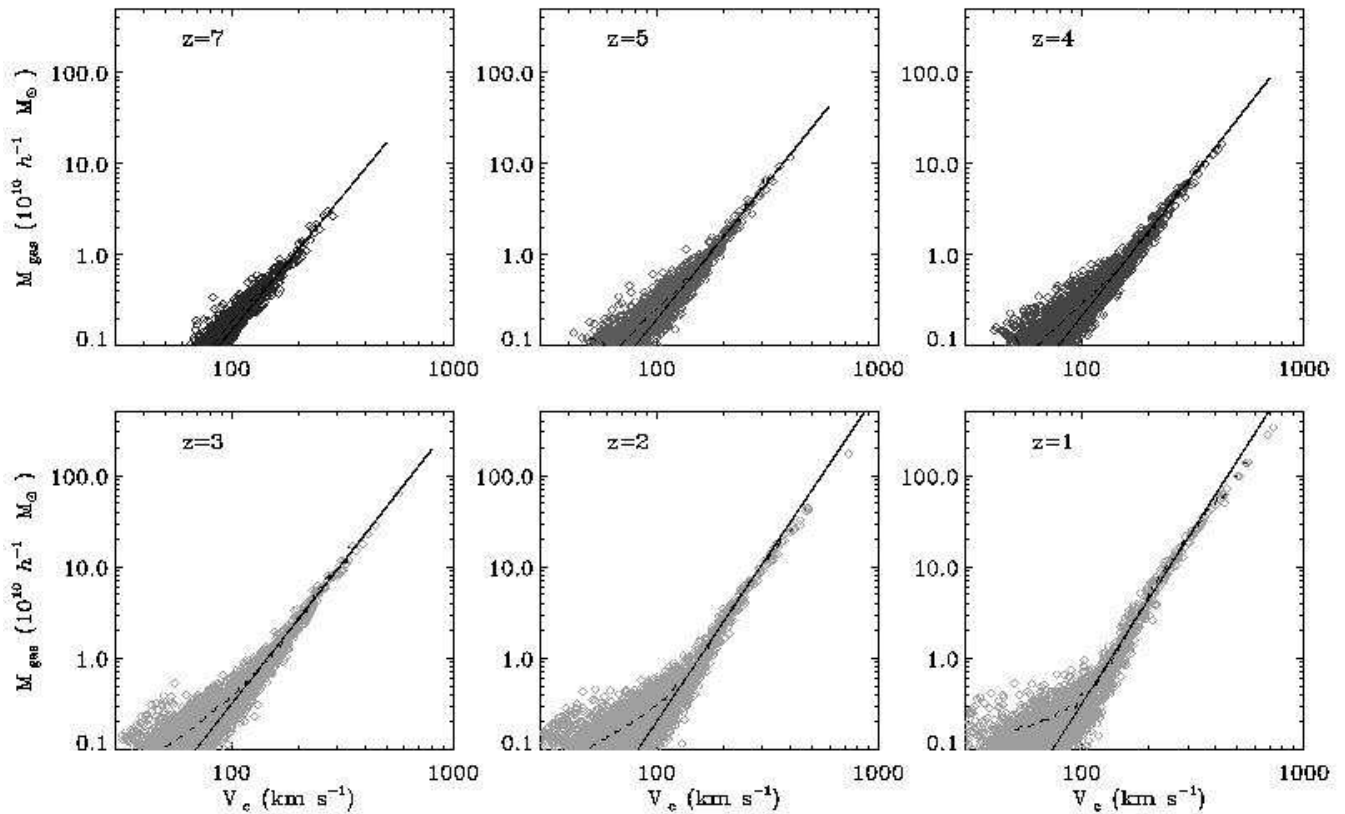,width=18.0truecm}}
}}
\caption{The evolution from $z=7$ to $z=1$ of total gas mass,
$M_{\rm gas}$, versus circular velocity $V_{c}$, for galaxies in the
D5 simulation. The dashed line shows the mean relation. The
solid line shows a power-law (in log) fit to the points. The power-law
index of the fitted line varies from 2.95 at $z=7$ to 3.98 at $z=1$ (see text).}
\label{fig_mgasz}	
\end{figure*}

\subsection{Galaxy definition}
For the galaxy selection we used the Friends-of-Friends (FoF)
algorithm to find groups of particles in the various simulation
outputs. A linking length of 0.1 times the mean spacing of dark matter
particles was used (also equal to $0.1\times 2^{1/3}=0.125$
 times the mean spacing of the
initial total number of particles). The algorithm was applied to all
star, gas and dark matter particles.  In choosing the linking length,
we have ignored the increase in the total number of particles due to
creation of star particles as the simulation evolves. Making
allowances for this would change the linking length by about $1.5\%$
in the most extreme case. We discarded groups that contained fewer than
32 particles and also those with either no gas or no star particles. The
FoF algorithm was chosen in the interests of simplicity, but we have
checked that using another groupfinder such as SKID (e.g., Governato
et al. 1997) makes no significant difference to our results.

\begin{figure}[t]
\hspace{2cm}        
\centerline{\hspace{3cm}\epsfysize=3.75in\epsffile{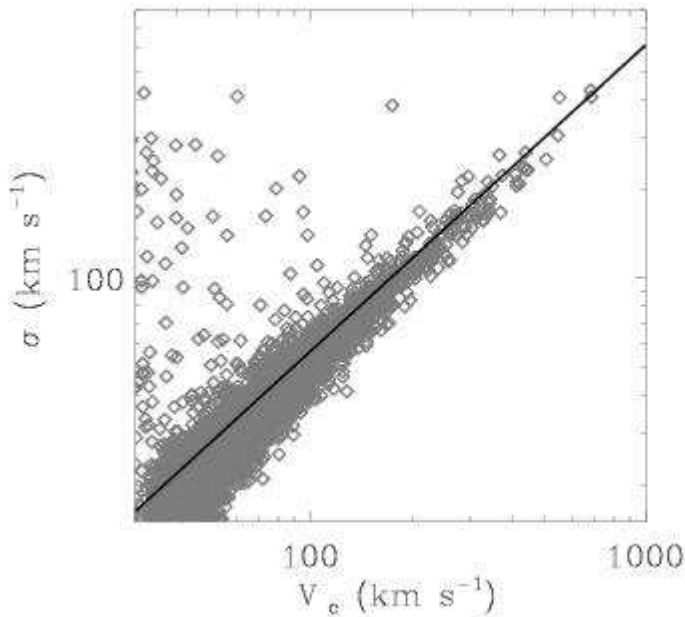}}
\caption {\label{figdisp} The 3D velocity dispersion of star
particles, $\sigma$, versus circular velocity, $V_{c}$, of galaxies
in the D5 simulation output at $z=1$. The solid line is a linear fit
(in log space) to the points.}
\end{figure}

\begin{figure}[b]
\centerline{
\hspace{2cm}
\psfig{file=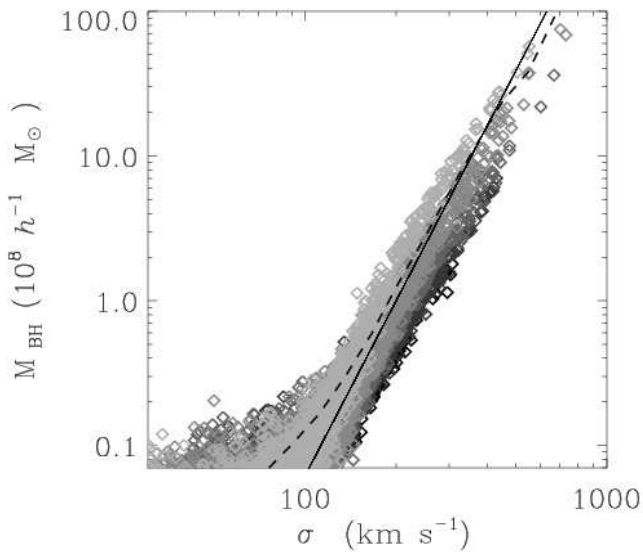,width=12.0truecm}
}
\caption{\label{figMbhSigma}The relation we obtain between
  $M_{\rm BH}$ (assumed to be proportional to the total gas mass) and
  velocity dispersion $\sigma$ of stars for groups in the `D5'
  simulation. The dashed line shows the mean relation over all
  redshifts, i.e.~for $z=7$, $6$, $5$, $4$, $3.5$, $3$, $2.5$, $2$,
  $1.5$, and $1$. For comparison, the solid line shows a power-law
  with a logarithmic slope of 4.}
\end{figure}

For each redshift output, we obtain a list of halos with their stellar
($M_{\rm star}$), baryonic ($M_b = M_{\rm gas} + M_{\rm star}$) and
dark matter ($M_{\rm DM}$) mass components, and their positions. The
total mass ($M = M_{b} + M_{\rm DM}$) of each galaxy is used to give a
measure of its radius and circular velocity. Following Mo \& White
(2002) and Springel \& Hernquist (2003b), we assign a physical radius
to a group of mass $M$ according to
\beq
r_{500} = \left[ \frac{GM}{250 \Omega_{m}(z) H^2(z)} \right] = 
\frac{1}{1+z} \left[\frac{GM}{250 \Omega_0 H_{0}^2} \right]^{1/3},
\eeq
and a corresponding circular velocity $V_{c} = (GM/r_{500})^{1/2}$,
which may be expressed as
\beq
V_{c} = (1+z)^{1/2} (GM)^{1/3} (250 \Omega_0 H_{0}^2)^{1/6}.
\label{eqn_vc_M}
\eeq

In these equations, we assumed that the selected groups on average
enclose an overdensity of 500 times the background value, which is
slightly higher than the canonical value of 200 usually adopted in
dark matter only simulations when a linking length of 0.2 is employed.
In this way, we compensate for the bias in the measured group masses
that arises from the slightly different distributions of gas and dark
matter particles on the scale of the `virial radius' $r_{200}$. Within
this radius, the gas particles are actually slightly underabundant
compared to dark matter particles, such that we effectively select
groups at a somewhat higher overdensity contour compared to a
corresponding dark matter only simulation. In fact, we have directly
compared our measured group masses to an alternative selection that
defines groups based on FoF applied to the dark matter only, using the
canonical linking length of 0.2, and increasing the final group mass
by the enclosed gas and star particles. In this case, the density
contrast with respect to the background can be assumed to be 200, and
the group mass $M_{200}$ may be defined to represent the `virial
mass'. We have correlated the two mass definitions using an
object-to-object comparison and found very small scatter between the
two different selection methods, with the two masses being related by
\beq
M_{200} \simeq 1.6 M,
\eeq
which can be used to approximately convert the masses we quote in this
paper to the virial masses of halos. We will also use the $M_{200}$
catalogues to compare the differences between the baryon fraction in
objects selected with the two different methods (see \S3.2).

For each galaxy we also directly calculate 3D stellar velocity
dispersions, $\sigma$, from the star particles. Finally, we measure
the amount of cold gas $M_{\rm cold}$ in galaxies, which we here
define to be the mass of actively star-forming cold clouds in the ISM.

\section{Cosmological black hole growth }
In Figure~1, we plot the gas mass $M_{\rm gas}$ for each individual
object selected in the six simulation outputs at redshifts $z=7$, $5$,
$4$ , $3$, $2$, and $1$, versus its circular velocity $V_{c}$. The
dashed line shows the average relationship traced by the galaxies at
the different redshifts and the solid line represents the best 
power law fit to the objects with $V_{c} \ge 80 \kmps$.
The fit has functional form
\beq
\log \left(\frac{M_{\rm gas}}{10^{10} h^{-1} \Msun}\right) = a \log \left(\frac{V_{c}}{200 \kmps}\right) + b,
\label{eqn_Mgas_vc}
\eeq
where the constants $a$ and $b$ for the six redshifts are $(z=7; 2.95,
-9.02)$; $(z=5; 3.00, -9.02)$; $(z=4; 3.08, -9.12)$; $(z=3; 3.10,
-8.98)$; $(z=2; 3.65, -10.2)$; $(z=1; 3.98, -10.4)$.
 Here we do not
attempt to assess the significance of the correlations as the sources
of scatter in the measurement of $M_{\rm gas}$ cannot be easily
quantified (see \S 3.1); the fitting is merely intended to illustrate
the overall evolution of the slope, $a$, that best describes the
relation. The most remarkable feature that we note is the steepening
of the $M_{\rm gas}-V_{c}$ relation from values $a\sim 3$ at high
redshifts, to $a\sim 4$ at $z \approxlt 2$.

We then compare the slopes (in log) of the $M_{\rm gas}-V_{c}$
relation to the most
recently revised $M_{\rm BH}-\sigma$ relation from Tremaine et al. (2002)
\beq
M_{\rm BH}=(1.5 \pm 0.2) \times 10^{8} \Msun \left(\frac{\sigma}{200
\kmps}\right)^{4.02 \pm 0.32}.
\label{eqn_Mbh_sigmaobs}
\eeq 
(Note that the first published estimates of the slope (in log) of the
above relation varied from values $\sim 4.8$; Ferrarese \& Merritt
(2000); Merritt \& Ferrarese (2001); to $\sim 3.75$ Gebhardt et
al. (2000), highlighting the difficulty of these measurements).  In
our lowest redshift output, at $z=1$, the observed slope of the
$M_{\rm BH} -\sigma$ relation (Eq. \ref{eqn_Mbh_sigmaobs}) is
reproduced well by the $M_{\rm gas}-V_{c}$ relation (Equation
\ref{eqn_Mgas_vc}), suggesting that the slope of Equation
\ref{eqn_Mbh_sigmaobs} is easily explained if $M_{\rm BH}$ is
directly proportional to $M_{\rm gas}$ (similar slopes are
obtained if we restrict ourselves to $V_{c} \approxlt 400 \kmps$,
 consistent with the observed velocity range  used to derive
 the Tremaine et al. 2002 relation). At
increasingly higher redshifts, the $M_{\rm gas}-V_{c}$ relation
flattens and approaches $M_{\rm gas} \propto V_{c}^3$ at $z\sim 7$, as
shown by the fitted slopes in Equation~(\ref{eqn_Mgas_vc}) and as
expected from Equation (\ref{eqn_vc_M}).

To make more direct contact with the observed $M_{\rm BH} -\sigma$
relation and to consider its possible link with the $M_{\rm
gas}-V_{c}$ relation in the simulations, we need to consider the
relationship between $V_{c}$ and the stellar velocity dispersion
$\sigma$, as measured directly from the simulations. 
The results for the $\sigma-V_{c}$ relationship measured in the
simulation are shown in Figure~\ref{figdisp}; the plot
shows the $z=1$ simulation output, although we note that
we find no significant temporal evolution in this relation. 
There is a direct proportionality between $V_{c}$ and
$\sigma$,\symbolfootnote[2]{The direct proportionality between $V_{c}$
and $\sigma$ also suggests that our group definition methods described
in \S 2.1 should be a fairly reliable in their identification of galaxies.
Note that the scattering of points above the mean relation in
Figure~\ref{figdisp} is most likely due to merging systems.}  which
we fit with the linear relation (in log):
\beq
\log \;\sigma = 1.03 \log \; V_{c} - 0.32,
\label{eqn_vc_sigma}
\eeq
shown by the solid line in Figure ~\ref{figdisp}. Because
the measurement of $\sigma$ from the star particles is intrinsically much
noisier than that of $V_{c}$ (which is effectively a measure of
mass), we make use of the relation from (Eq.\ref{eqn_vc_sigma})
to convert from $V_{c}$ to $\sigma$ in what follows.  
Using $\sigma$ directly would introduce a significant source
of artificial scatter in our simulated $M_{\rm BH} -\sigma$
relation. We note that in future work it might be possible to
investigate the scatter about the $M_{\rm BH} -\sigma$ relation
provided it is possible to eliminate numerical scatter in the
simulated stellar velocity dispersion $\sigma$.

The two relations, $M_{\rm gas}-V_{c}$ and $V_{c} - \sigma$ from the
simulations, if compared to the observed $M_{\rm BH}-\sigma$ relation
(Eq.~\ref{eqn_Mbh_sigmaobs}) suggest that the central black hole mass in
galaxy may be simply proportional to the gas mass in galaxies.
We hence make this simple ansatz that the black hole
mass is proportional to the gas mass, allowing us to plot a predicted
relation between $M_{\rm BH}$ and $\sigma$ obtained from the
simulations.  In Figure~\ref{figMbhSigma}, we overplot the
corresponding data from all the redshift outputs used in Figure 1,
resulting in a composite average relation which is shown as a dashed
line. Only simulation outputs for the D5 run are included, and we
defer a discussion of differences between the D4 and D5 runs to the
next section. In order to match the normalization of the observed
$M_{\rm BH}-\sigma$ relation (Eq.~\ref{eqn_Mbh_sigmaobs})
we associate black hole and gas masses here according
to
\beq
M_{\rm BH} = f \; M_{\rm gas} \sim 0.004 h^{-1} M_{\rm gas}, 
\label{eqn_Mgas_MBH}
\eeq
where the solid line in Figure~\ref{figMbhSigma} shows
Equation~\ref{eqn_Mbh_sigmaobs}. Assuming that there is no significant
change from $z=1$ to $z=0$, (which we have checked is a good
approximation by making use of the larger 100 h$^{-1}$ Mpc simulation box in
the G-Series; see Springel \& Hernquist 2003a) this therefore implies
that in the context of our model, the mass of the central black hole 
in galaxies is consistent with being
$\approxlt 1$ per cent of the total gas mass in galaxies.

In this interpretation, the $M_{\rm BH} -\sigma$ relation is not set
in primordial structures and maintained through cosmic time; it is
established at later times as the growth of black holes is limited by
the available gas, in competition with star formation and associated
feedback processes like galactic winds. Because both the black hole
growth and the change of the amount of available gas occur gradually
from high to low redshifts, the resulting correlation does not show
large scatter. Note, however, that in this simple model, the high-mass,
high-redshift objects are predicted to lie on the right of the main
relation. Large samples of intermediate mass objects at $z \approxgt
3$ will be required to determine observationally whether the $M_{\rm
BH} -\sigma$ relation 
is indeed shallower at earlier times, as predicted here.
At present the observed $M_{\rm BH} -\sigma$ relation (as shown in
Eq.~\ref{eqn_Mbh_sigmaobs}) is derived from local objects. 
For objects at $z=1$ (sixth panel of Fig.~\ref{fig_mgasz})
the predicted relation is extremely tight.

Note also that given the $\sigma - V_{c}$ relation 
(Eqs.~\ref{eqn_vc_sigma} and ~\ref{eqn_Mgas_MBH}) from the
simulations we find that the black hole mass is related to the dark
matter mass of the host galaxy by the relation:
\beq
\frac{M_{\rm BH}}{10^8 \Msun} \sim 0.7 
\left(\frac{M_{\rm DM}}{10^{12} \Msun}\right)^{4/3}.  
\eeq
This is broadly consistent with the $M_{\rm BH} - V_{c}$ relation
derived by Ferrarese (2002) and Baes et al. (2003),
reflecting  the fact that the relation in
Equation~\ref{eqn_vc_sigma} matches the observed correlation between
$V_{c}$ and $\sigma$ (Ferrarese 2002; Baes et al. 2003).

\subsection{Numerical convergence and scatter}

\subsubsection{Deviations from a power law}

Figures 1 and \ref{figMbhSigma} show that, for small values of $M_{\rm
gas}$ and $V_{c}$, the $M_{\rm BH}-\sigma$ relation deviates
significantly from a power-law and its scatter is increased. Figure 4
shows the same plot as in Figure 1 but with the D3, D4 and D5
simulations over-plotted at $z=1$ in the top panel, and the D5 and Q5
simulations in the bottom panel. The `Q5' simulation is taken from the
Q-Series of Springel \& Hernquist (2003b). While Q5 has a smaller
cosmological volume of size 10$h^{-1}$ Mpc on a side, its particle
number is equal to that of D5, so that the mass resolution in the gas
becomes $3.26 \times 10^5 h^{-1} \Msun$ (a factor of 40 improvement
over D5), and the spatial resolution becomes 0.63 $h^{-1} \kpc$ in
comoving units.

In the top panel of Figure~4, the average relations for the D3, D4 and
D5 runs are shown with the dashed, solid and dot-dashed lines,
respectively. The figure shows that both the scatter and the deviation
from the $V_{c}^4$ relation decrease with increasing resolution of the
simulation. This is a result of the fact that close to the resolution
limit of each of the simulations, star formation cannot be resolved
properly any more, making the measurement of $M_{\rm gas}$
unreliable. For higher resolution simulations, this limit for
convergence in $M_{\rm gas}$ moves to lower mass systems.

However, when analyzed in isolation, even for the D5 run it is not
clear whether the flattening of the relation at small $V_{c}$ is due
to some underlying physical process associated with star formation, or
whether it can be fully explained by insufficient resolution. We
investigate this further in the bottom panel of Figure 4, where we
show a comparison of the measured gas content between the D5 and the
Q5 simulations at redshift $z=3$ (this is close to the minimum redshift
reached by the Q-Series). Although the number of objects in the high
mass range is much reduced in Q5 due its smaller volume, the increased
resolution allows us to investigate the $M_{\rm gas}-V_{c}$ relation
for much smaller objects. We find almost negligible deviations from the
extrapolated power law behavior derived from the D-Series.  It is
therefore plausible that the $M_{\rm BH}-M_{\rm gas}$ relationship
extends to very low $V_{c}$ given sufficient numerical resolution.

\begin{figure}[t]
\center{
\vbox{
\psfig{file=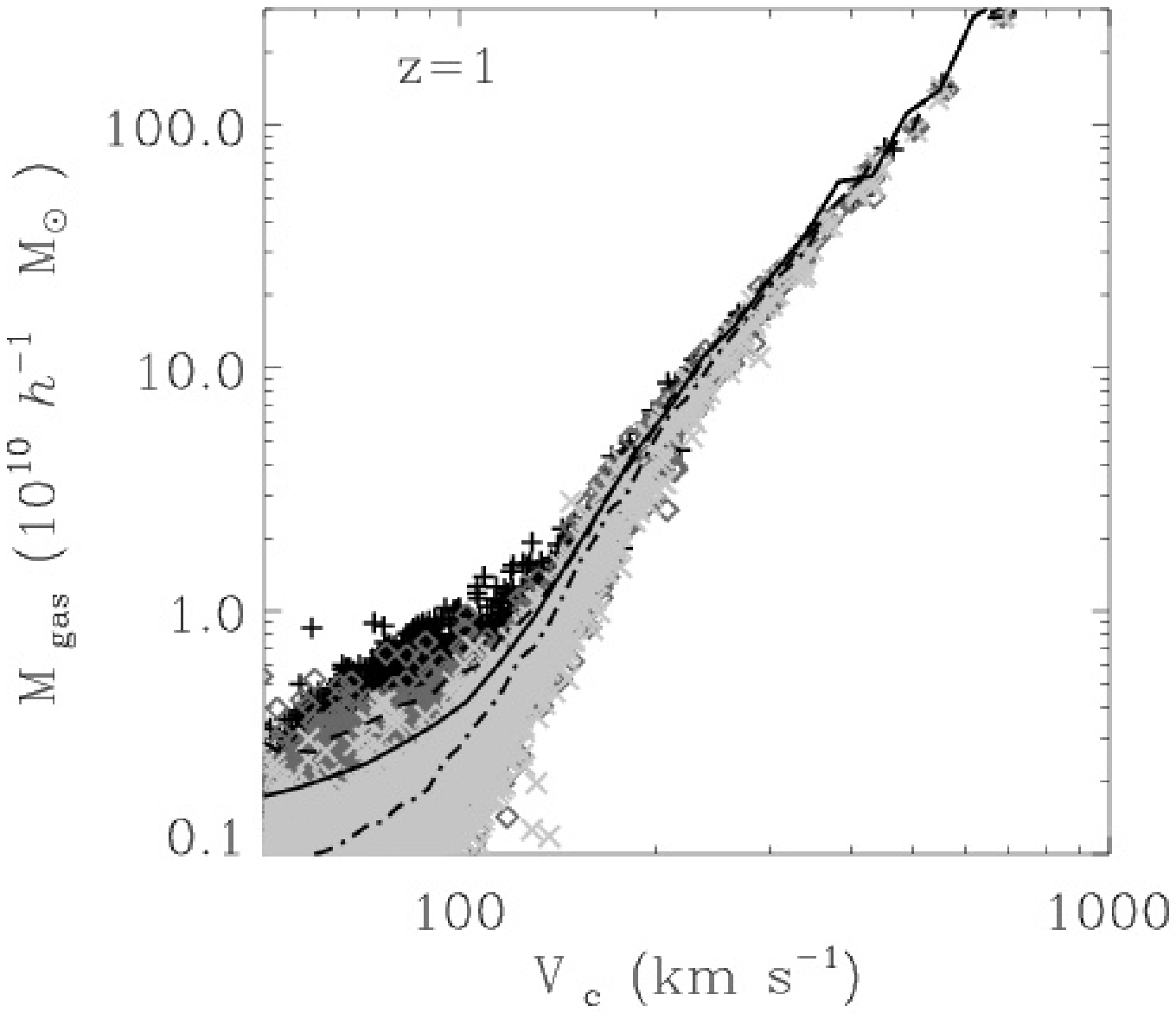,width=12cm}
\psfig{file=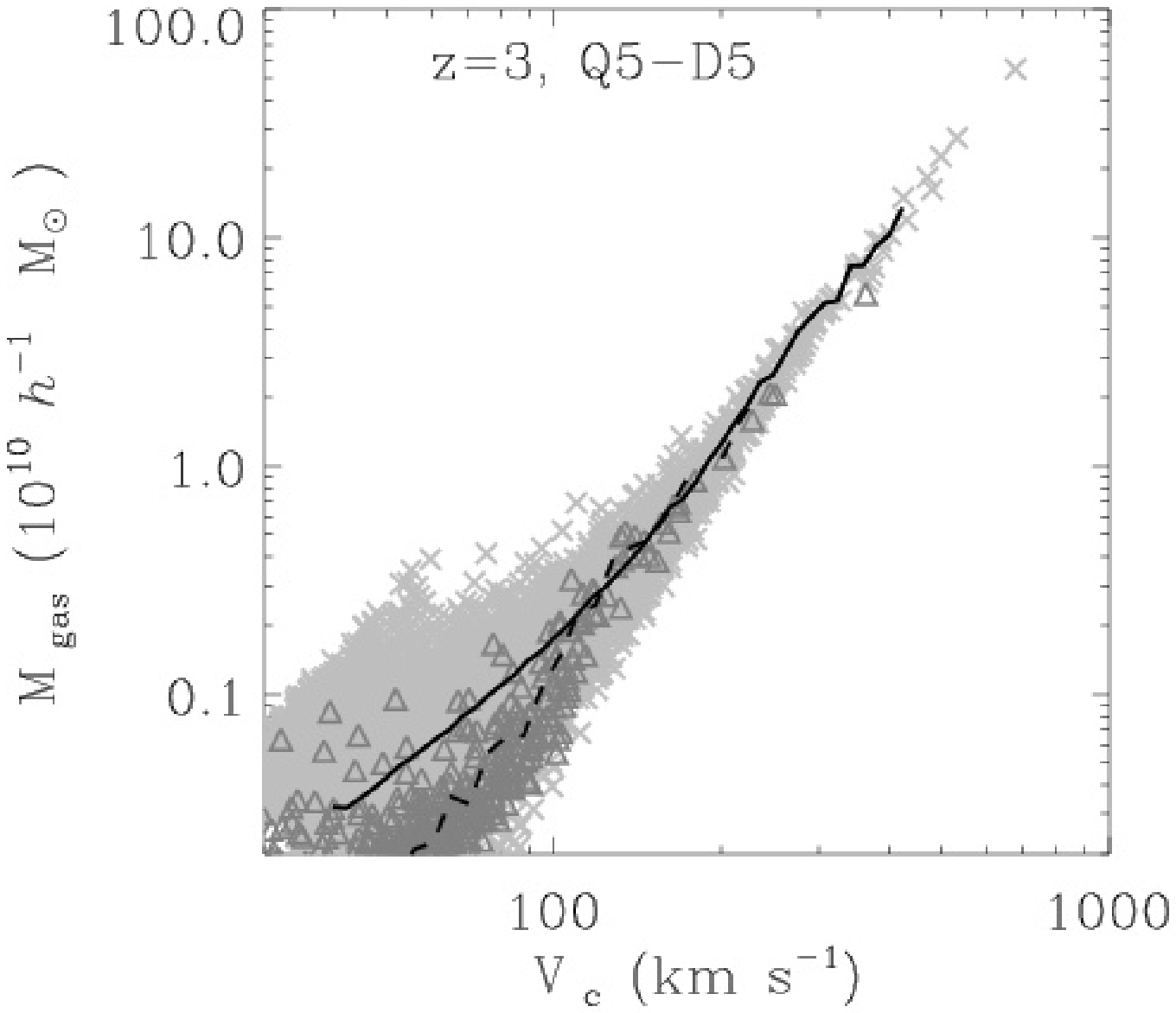,width=12cm}}}
\caption{Convergence study for the measurement of $M_{\rm gas}$. We
plot the same quantities as in Figure 1. Top panel: groups extracted
from the D3, D4 and D5 simulations at redshift $z=1$ are overplotted
with plus signs, triangles and cross symbols, and in shades from dark
to light gray, respectively. The dashed line is the average relation
in the D3 run, the solid line in D4 and the dash-dotted in D5. The
large scatter at small $V_{c}$ is reduced with increased resolution.
Bottom panel: We compare here objects in the Q5 (dashed line)
 and D5 simulations at
redshift $z=3$. Thanks to the significantly increased resolution of
the Q5 simulation, the scatter is further reduced, showing that it
largely arises from resolution effects. However, the small box of Q5
contains only a few high-mass objects.}
\end{figure} 

Determining the extent to which our theoretical relation applies to
the low mass range is also relevant in view of recent searches for
intermediate mass black holes in globular clusters. The promising
detection of a central black hole of $M_{\rm BH} \sim 2\times 10^{3}
\Msun $ in stellar cluster G1 in M31 (Gebhardt, Rich \& Ho, 2002) and
the ongoing debate on the presence of a similar central object in the
globular cluster M15 (Gerssen et al. 2002) seem to be consistent
with an extrapolation of the observed $M_{\rm BH}-\sigma$ relation for
galaxies, possibly implying a similar formation process for the
central black hole.

Figures 1 and \ref{figMbhSigma} also show that for large values of
$V_{c}$ the relationship between $M_{\rm gas}$ and $V_{c}$ flattens
and almost recovers the power law slope of $\sim 3$ expected from
Equation  (\ref{eqn_vc_M}). for halos with baryon content consistent with the
universal fraction. As we will
discuss further in \S 3.2, stellar feedback with its associated
outflows plays a fundamental role in our model for generating the
steep decline of gas content towards low-mass halos, but these
processes eventually become inefficient for very massive systems,
which then recover the universal baryon content.

\subsection{The evolution of 
$M_{\rm gas}$, $M_{\rm star}$ and $M_{\rm cold}$}

In order to establish the constant of proportionality $f$ in
Equation~(\ref{eqn_Mgas_MBH}) we need to investigate whether the
measurement of $M_{\rm gas}$ has converged for sufficiently large
$V_{c}\approxgt 100 \kmps$.  Figure 5 shows the evolution of the total
gas, stellar and cold gas (gas subject to star formation) masses, $M_{\rm
gas}$, $M_{\rm star}$ and $M_{\rm cold}$ as a function of $z$ at
constant circular velocity, $V_{c} = 200\kmps$. We also include the
dark matter component, $M_{\rm DM}$, for comparison. The plot hence
shows the evolution of the baryonic (and dark matter) content in
typical $L_\star$-galaxies. The black symbols and lines are for the D3
simulation, the light gray for the D4 run and the dark gray for the D5
run. For each simulation, the star symbols joined by dot-dashed lines
represent $M_{\rm star}$, the triangles joined by the dashed lines
$M_{\rm cold}$, and the solid dots joined by the solid lines $M_{\rm
gas}$.  $M_{\rm DM}$ is shown for the D5 run only, using diamonds
connected by a dotted line and rescaled by a factor
$\Omega_{b}/\Omega_{\rm DM} = 0.153$.

The figure shows that the measurements of $M_{\rm gas}$, $M_{\rm
star}$ and $M_{\rm cold}$ at $V_{c} = 200 \kmps$ have converged in the
D4 run to good accuracy. In the D3 simulation, the star formation has
not fully converged even at this circular velocity; hence $M_{\rm
star}$ is lower in the D3 run than in D4 or D5, and $M_{\rm cold}$, in
particular, as well as $M_{\rm gas}$ are overestimated.

It is further seen that the ratio $M_{\rm gas}/M_{\rm star}$ is large
at high redshifts and approaches values $\approxgt 1$ at redshifts $z
\approxlt 3$. The cold gas fraction $M_{\rm cold}/M_{\rm star}$ is
also fairly large at high redshift but decreases significantly below
$1$ at $z\sim 1$.  Interestingly, in this plot we see the overall
evolution of the baryon and dark matter mass in galaxies. This will be
particularly useful when deriving the evolution of the quasar
luminosity function in \S4.  Here we find that $M_{\rm gas}$ for the
circular velocity given, $V_{c} = 200\kmps$, is a slowly decreasing
function of redshift, where we have $M_{\rm gas}\propto (1+z)
^{-1/3}$, as opposed to $M_{\rm DM}$, which is a steeper function of
redshift, following the relation $M_{\rm DM} \propto (1+z)^{-3/2}$ as
expected from Equation  (\ref{eqn_vc_M}). Even though 
gas is both used up by star formation and expelled by galactic
outflows, the total mass of gas does not decrease with decreasing
redshift but remains nearly constant.

Note that in the Kauffman \& Haehnelt (1999) model, in order to explain the
small scatter in the $M_{\rm BH}$ - $\sigma$ relation, Haehnelt \&
Kauffmann (2000) require the cold gas mass of their bulge progenitors
to be essentially independent of redshift (where in their model a
fixed fraction of the cold gas is assumed to to be accreted by the
black holes). Here we find that $M_{\rm cold}$ decreases with redshift
and instead the total gas mass $M_{\rm gas}$ is found to be a slowly
varying function of redshift. This suggests that the slope of the
$M_{\rm BH}$ - $\sigma$ relationship may not be set by the amount of
gas that cools and becomes available for star formation, as assumed by
Haehnelt \& Kauffmann (2000), but rather by the total amount of gas in
the galaxies, although it is certainly true that galaxies have higher
cold gas fractions at earlier times. In the next section we
investigate how star formation and feedback regulate the total amount
of gas available for accretion onto central black holes.

\begin{figure}[t]
\hspace{2cm}        
\centerline{\hspace{3cm}\epsfysize=3.75in\epsffile{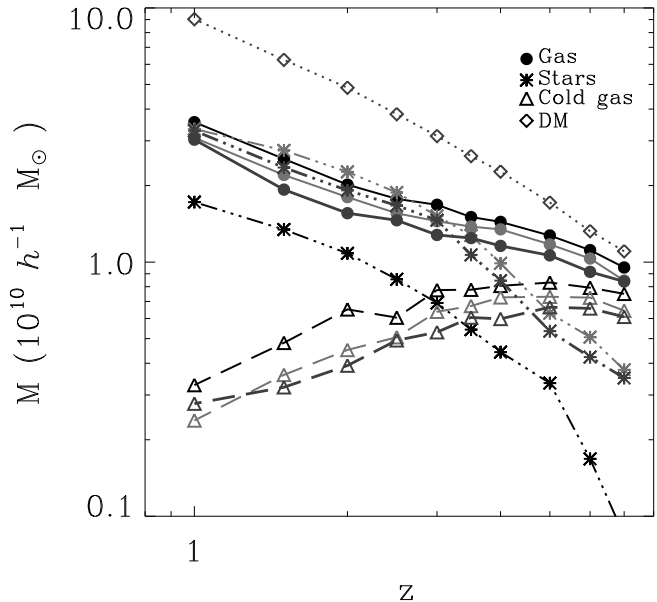}}
\caption{Gas, star, cold gas and dark matter mass components for halos
of constant circular velocity $V_{c} = 200 \kmps$ as a function of
redshift. The results for the D3 run are shown in black, for D4 in
light gray, and for D5 in dark gray.  The dark matter mass has been
rescaled by a factor $\Omega_{b}/\Omega_{\rm DM} = 0.153$ and is shown
for comparison only (for D5). Approximate convergence of the various
baryonic mass components is already obtained for the resolution of the
D4 simulation.}
\label{fig:m_evol}
\end{figure}

\begin{figure}[t]
\center{
\vbox{
\psfig{file=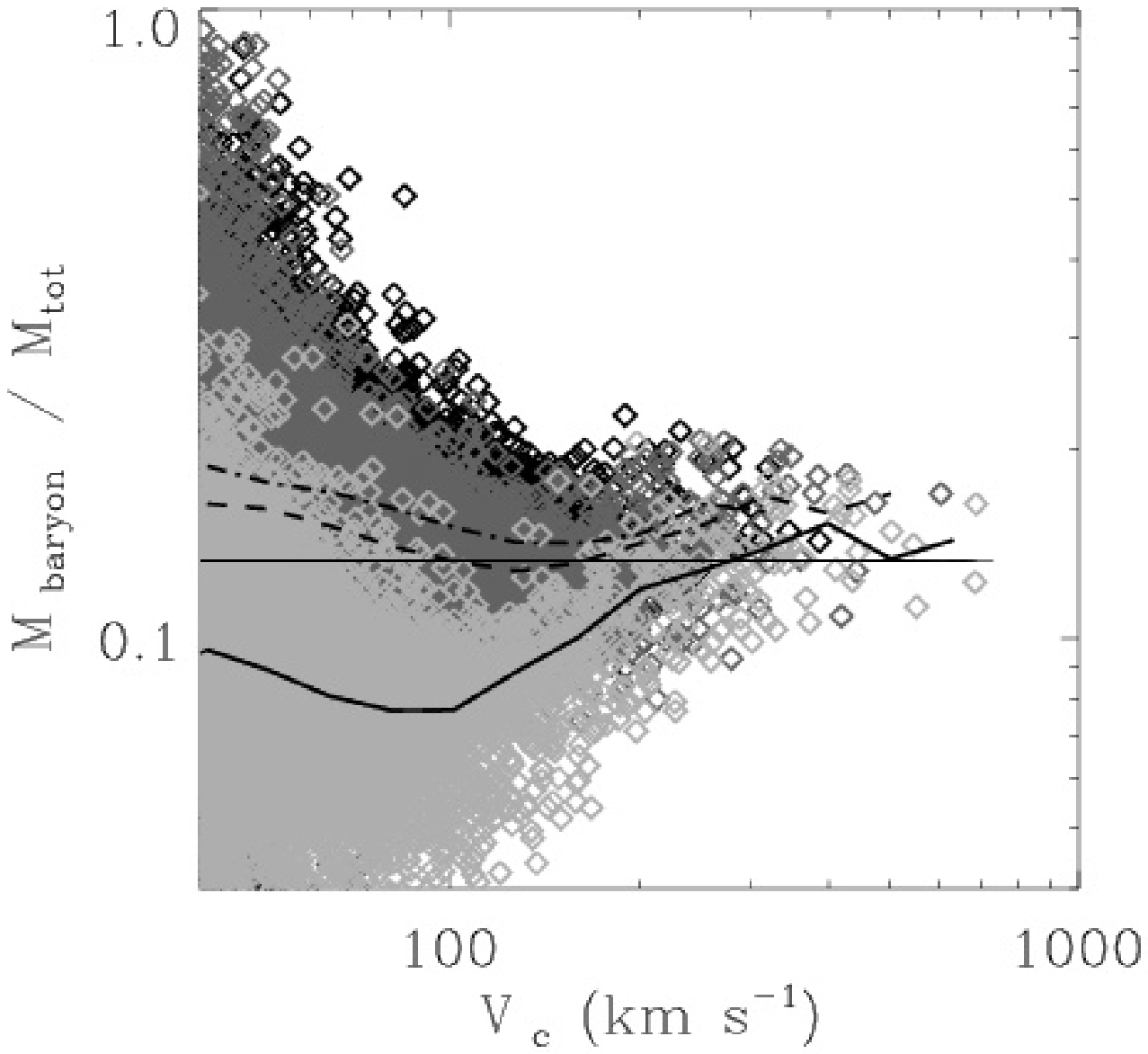,width=12cm}
\vspace{-1cm}
\psfig{file=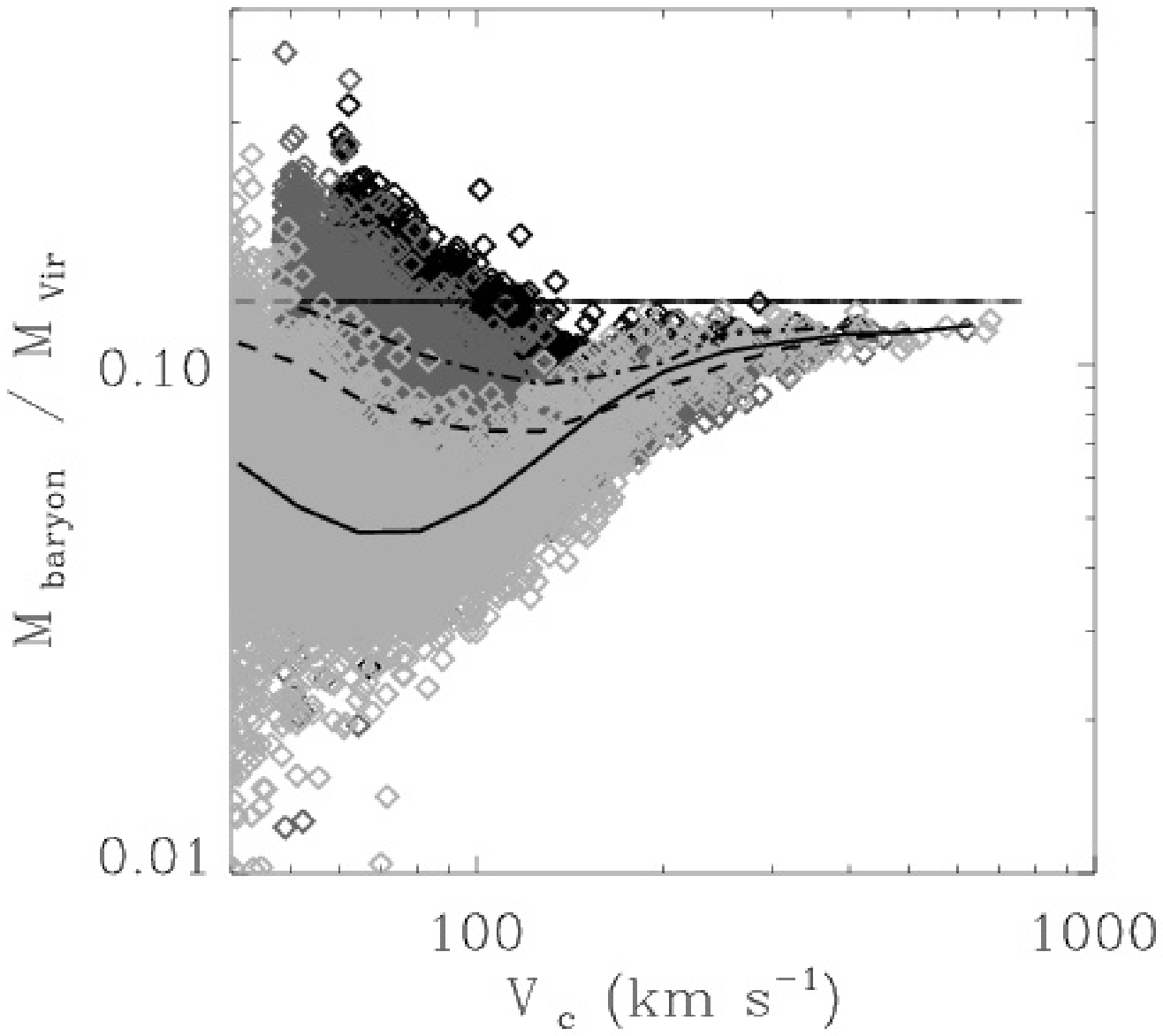,width=12cm}}}
\caption{The baryon mass fraction in galaxies as a function of
circular velocity $V_{\rm c}$. Three redshifts, $z=5$, $3$, and $1$,
are shown in shades of dark to light gray, and we plot running
averages for them in dot-dashed, dashed and solid lines,
respectively. The horizontal line shows the baryon fraction
$\Omega_{\rm b}/\Omega_{\rm m}$.  In the top panel, objects were
selected by applying a FoF linking length of 0.1 the initial dark
matter particle spacing to all particles, with the total mass in
galaxies defined as $M_{\rm tot} = M_{\rm gas} + M_{\rm stars} +
M_{\rm DM}$. In the bottom panel, a linking length of 0.2 has been
used instead, but was restricted to the set of dark matter
particles. Here, the enclosed gas and star particles in each
identified group have been added in a separate step (see~\S~2.1).}
\end{figure}

\subsection{The role of stellar feedback}

In Figure 6, we plot the baryonic mass fraction $(M_{\rm gas} + M_{\rm
star}) / M_{\rm tot}$ in galaxies as a function of $V_{c}$ for the
objects in the D5 simulation at the three redshifts $z=5$, $3$, and
$1$, overplotted in black, dark gray and light gray, respectively. A
running average of the baryon fraction at each redshift is shown as a
dash-dotted line for $z=5$, dashed line for $z=3$ and solid line for
$z=1$. The solid horizontal line indicates the universal baryon
fraction $\Omega_{b}/\Omega_{\rm m}$ expected for our cosmological
model.  The two panels of the figure show a comparison of the baryon
fractions measured for objects selected according to the two group
definitions outlined in \S 2.1. The top panel uses the same definition
as in our previous figures, where we applied the FoF algorithm to dark
matter, gas and star particles on an equal footing, while in the
bottom panel, we selected objects based on dark matter particles only,
and then included all enclosed gas and star particles.  Based on
this comparison, we note that the larger scatter in the top panel at
low $V_{c}$ is a consequence of the different group selection
methods. In particular, close to the resolution limit our default
method is somewhat more prone to occasionally producing measured baryon
fractions well above $\Omega_{b}/\Omega_{\rm m}$, as it can happen if
the FoF algorithm just picks out the concentrated cold gas and stellar
parts of low-mass galaxies, but loses the sparsely sampled
dark halo around them. Although this occurs, our method is able to recover
the universal baryon fraction at $z=1$ in the simulation more
closely (top panel, Fig. 6) than in the bottom panel.

However, more importantly, both plots show that the total baryon mass
fraction decreases from intermediate values of $V_{c}$ towards low
values, while systems of very large mass remain close to the universal
baryon fraction. Furthermore, the strength of the baryonic depletion
at low $V_{c}$ increases towards low redshifts.

It is clear that these trends in Figure 6 are a reflection of the same
steeping of the $M_{\rm gas}-V_{c}$ relation towards low redshift and
low $V_{c}$ that we already observed in Figure~1. However, in this
representation the fundamental physical mechanism that drives this
evolution is more apparent. In essence, the $M_{\rm gas} - V_{c}$
relation is steepened by the loss of baryons due to strong feedback by
galactic winds.  These galactic outflows are increasingly more
efficient in low mass systems (where resolved; see also Springel
\& Hernquist 2003b), while for very massive groups, the winds
eventually become trapped in deep gravitational potential wells.  Note
that the measured $M_{\rm gas}$-$V_{c}$ relation indeed flattens for
high-mass systems (as shown by Figures 1 and 3), with a transition
scale that is consistent with the maximum circular velocity 
expected to allow an escape of winds from halos in the simulations.

\section{Black hole activity}
Quasars have long been believed to be powered by the accretion of gas
onto supermassive black holes and have been known to evolve very
strongly. Their comoving space density increases by nearly two orders
of magnitude out to a peak at $z\sim 3$ (e.g., Boyle et
al. 2000). Their evolution at higher redshifts is more uncertain
although recent observations have begun to constrain the bright end of
the luminosity function out to $z \sim 5$ 
and individual quasars have been detected out to $z=6.28$
(Fan et al.~2001a,b). Here we
use our results from \S 3 to predict the evolution of the number
density of bright QSOs from redshift $z=7$ to $z=1$ and compare it
with observations.

\subsection{Gas accretion history}

\begin{figure*}[t]
\center{
\centerline{\psfig{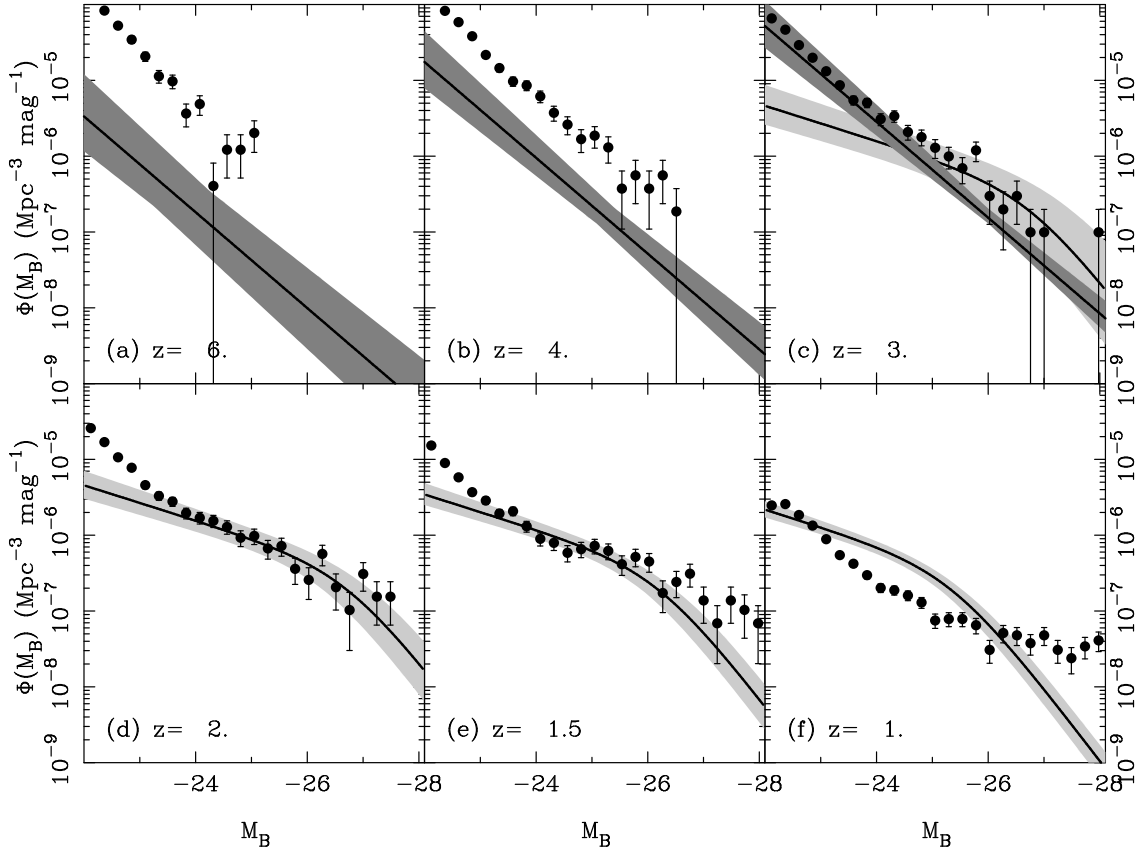}}}
\caption{The $B$-band luminosity function of quasars at various
redshifts, as indicated in each panel. The filled circles are the
values from our models. Error bars indicate the Poissonian error in
the counts. The solid lines and light gray shaded areas at $z=2$,
$1.5$, and $1$ show the best fit to the 2dF luminosity function (Boyle
et al. 2000). The dark shaded areas and solid lines at $z=6$, $4$, and
$3$ give the best fit to the SDSS luminosity function from Fan et
al.~(2001a,b).}
\end{figure*}

In order to predict black hole luminosities, we need to estimate the
amount of gas available for accretion for a given  characteristic
accretion timescale.  In \S 3 we showed that the mass of the black
hole may simply be modeled as a fixed fraction of the total amount of
gas in the halo, suggesting that black holes grow primarily by
accretion of gas, in which case the amount of accreted mass also
depends on star formation and feedback processes.

The small-scale physical processes linked to the growth of black
holes, however, do not give us a direct handle for constraining the
black hole accretion timescale. Given the lack of any definite
prescription or constraint for the quasar lifetime or its evolution,
we make the simplest possible assumption and choose the quasar
lifetime which is independent of redshift and of order of the Salpeter
time $t_{Q} \sim 2 \times 10^7$ yr. Our choice of $t_{Q}$ is most
appropriate for the bright end of the quasar population (for quasars
that radiate around the Eddington limit) which is what we concentrate
most of our modeling on, as the faint end is dominated by low mass
galaxies which may not be fully resolved in the simulations used here
(see \S3). We note that this choice of timescale is consistent with
results of Steidel et al. (2002) who estimated the lifetime of bright
quasar activity in a large sample of Lyman-break galaxies at $z \sim
3$ to be $\sim 10^7\,{\rm yr}$ and range of quasar lifetimes obtained
from quasar clustering constraints ( which generally imply $10^7 {\rm
yr} \approxlt t_{Q} \approxlt 10^8 {\rm yr}$; e.g.; Martini \&
Weinberg 2001). It is interesting to note that a similar timescale,
$t_{Q} \sim 2 \times 10^7$ yr, was also adopted in the models of
Sokasian, Abel and Hernquist (2002; 2003) developed to study
reionization of HeII and the properties of the ionizing background
at intermediate redshifts. In this case, this timescale was shown to
match the observations of the HeII Lyman alpha forest and of the H
and HeII opacity at $z \sim 2.5 - 5$.

As we shall see the choice of this parameter is crucial for
determining the evolution of the quasar population.  For our model, the
star formation timescale is $t_{\rm star} \sim$ a few Gyr; hence, we
have $t_{Q} << t_{\rm star} << H_0^{-1}$, consistent with our
hypothesis that the overall cosmological change in the environment
governs the quasar activity and the duty cycle of quasars increases
with redshift.

We assume that all galaxies can undergo a quasar phase and so
accrete a fraction of their total gas content. Because, in
general, the evolution of the gas mass in galaxies follows an
evolution $\approxpropto (1+z)^{1/3}$ (i.e.; 'on average' the gas mass
does not go down, but merely does not grow; see
Fig.~\ref{fig:m_evol}), we can assume the black hole growth 
occurring in the last Hubble time always dominates. Hence, we
consider the black hole mass at the time of accretion to be the
contemporaneous black hole mass.  This simple scenario is then consistent
with the approximate condition, $M_{\rm BH} \propto M_{\rm gas}$ relating
the total gas and black hole masses.

The bolometric luminosity of a quasar at redshift $z$ is then
simply given by
\beq
L = \eta \Mdot(z) c^2 \sim \eta \frac {\Delta M_{\rm gas}}{\Delta t} c^2
\sim
\eta f M_{\rm gas}(z) c^2/ t_{Q},
\label{eqn:Lacc}
\eeq
where we have taken $ \Delta M_{\rm gas} (z) \sim f M_{\rm gas} (z)$
with $f=0.4\%$, as suggested by Equation~(\ref{eqn_Mgas_MBH});
i.e.~the average mass accreted is a constant fraction of the total gas
mass $M_{\rm gas}$. We adopt the standard value for accretion
radiative efficiency of $\eta = 10 \%$, and assume that every object
with a nonzero star formation rate (this discards small-mass objects
where star formation is likely not resolved) shines as a quasar for a
time $t_{Q}$. The expected number density of active quasars at each
redshift is a fraction $f(z_{i}) = t_{Q} \times
(z_{i+1}-z_{i})/[t_{H}(z_{i})-t_{H}(z_{i+1})]$ where $z_{i}$ and
$t_{H}(z_{i})$ are the redshift and Hubble time of simulation output
$i$, respectively.

\subsection{The evolution of the quasar luminosity function}
\subsubsection{Optical blue band}
We now compare our simple predictions for the luminosity function to
observational data. We take a fixed fraction $\epsilon_{B}$ of the
bolometric luminosity to be radiated in the blue band.  In the
$B$-band, the bolometric correction $\epsilon_{B}$, defined as $L =
\epsilon_{B} \nu L_{B}$, is about 11.8 (Elvis et al.~1994), where $\nu
L_{B}$ is the energy radiated at the central frequency of the $B$-band
per unit time per logarithmic interval of frequency. At the peak of
its lightcurve the quasar has a $B$-band magnitude
\beq
M_{B} = 37.05 - 2.5 \log ({\epsilon_B L}),
\eeq
where $L$ is in units of $\ergps$.~We place an upper limit on the
maximum allowed luminosity of a quasar so that $L$ does not exceed the
Eddington luminosity $L_{\rm Edd}$. Note that if accreted gas provides
the power, then clearly there cannot be steady fueling at rates that
generate $L > L_{\rm Edd}$. With the black hole masses given by
Equation~(\ref{eqn_Mgas_MBH}) a small percentage of objects radiate
above $L_{\rm Edd}$.

Figure 7 shows a comparison of our theoretical luminosity functions
(LFs) at different redshifts with the most recent determinations of the
quasar $B$-band LF from the 2dF QSO Redshift survey ($ 0.3 < z < 2.3$;
Boyle et al. 2000) and the Sloan Digital Sky Survey SDSS ($3.6 < z
\approxlt 6$; Fan et al.  2001a). For the 2dF, the luminosity function
for $-26 < M_B < -23$ is described by a double power law with a bright
end slope of $\beta_1 = -3.4$ and a faint end slope of $\beta =-1.5$
(Boyle et al. 2000; Pei 1995). The QSO luminosity function measured by
Fan et al. (measured for $-25 < M_{1450} < -28$; where $ M_{1450}$ is
the absolute AB magnitude; see Eq (5) in Fan et al. 2001a for the
relation between $M_B$ and $M_{1450}$) gives a flatter bright-end
slope ($\beta_1 = -2.5$). For the luminosity function at $z=3$, we
plot both these functions, one extrapolated from higher redshifts, the
other from lower redshifts.

Our simple model reproduces the luminosity function of optically
selected quasars in the redshift range $4 \approxlt z
\approxlt 1.5$ reasonably well. 
The overall turnover of the space density of quasar at low redshifts
($z \approxlt 3$) is also reproduced with some overestimate of the
number of bright quasars at $z=1$.  The slope of the LF at redshifts
higher than $z\approxgt 4$ matches the steep value found by Fan et
al. (2001a). However, the model at this high redshifts predicts a larger
numbers of quasars (both bright and faint) with respect to
observations (and their extrapolation to faint magnitudes $M_{B}
\approxgt -25$).
At redshifts $z = 3$ and $2$, the bright end of the luminosity
function flattens in accordance with the observations by Boyle et
al. (2000) but remains too flat at $z=1$, leading to an overestimate
of the number of bright quasars at these redshifts.

In the top panel of Figure 8, we show the evolution of the space
density of bright quasars with luminosity $M_{B} < -26$ as a function
of redshift.  In this plot, we add as open symbols data from the
Warren, Hewett, \& Osmer survey (1994; hereafter WHO) in the redshift
range $2 < z < 4.5$, and from Hartwick \& Shade (1990) in the range
$0.1 < z < 3.3$, respectively (also summarized in Pei 1995). The
dotted line is a fit to the Schmidt, Schneider \& Gunn survey (1995;
SSG). The Fan et al. (2001a,b) data is shown with solid dots, while
our theoretical prediction is shown using solid symbols joined by a
solid line. Although this measurement is quite noisy as a result of a
relatively small simulation box, the overall evolution of the comoving
number density of quasars is reproduced reasonably well by the
model. Note that because of our limited box size no objects can be found
massive enough to contribute to the luminosity function at $M_{B}
< -26$ for $z >5$. As we shall discuss in \S 4.2.2 , at high
redshifts, black hole accretion follows the growth of structure; to a
good approximation our prediction at higher redshifts should thus
follow the extrapolation from lower redshifts.  Because of the small
number of objects and the very sharp decrease in the quasar counts for
objects brighter than $-26$ our results are very sensitive to the
model parameter $t_{Q}$. In order to get a more robust comparison we
compare our results to recent constraints on quasar number counts
derived from recent Chandra X-ray observations which probe quasars of
more intermediate luminosity.

\begin{figure}[t]
\center{
\vbox{
\psfig{file=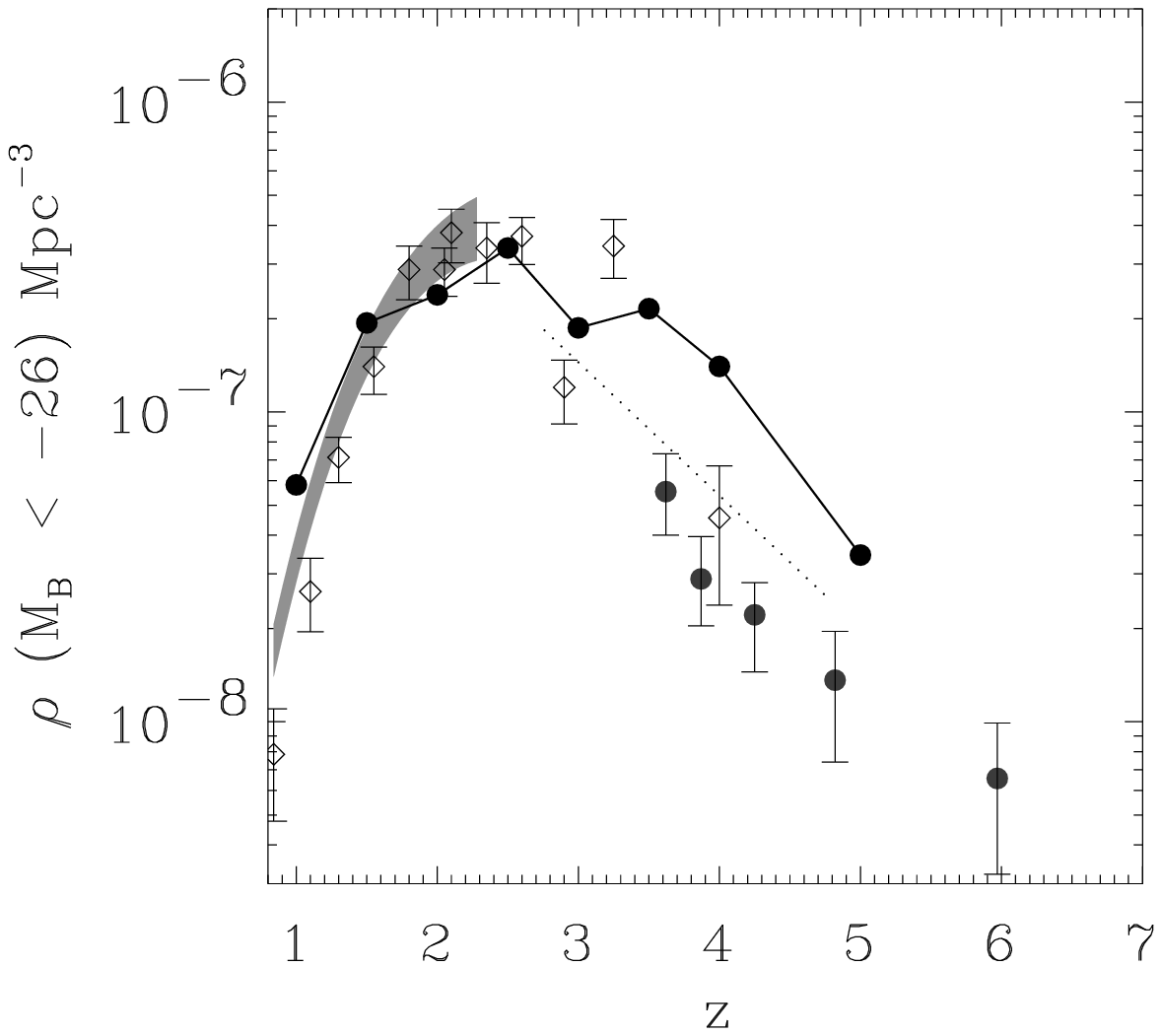,width=12.0truecm}
\psfig{file=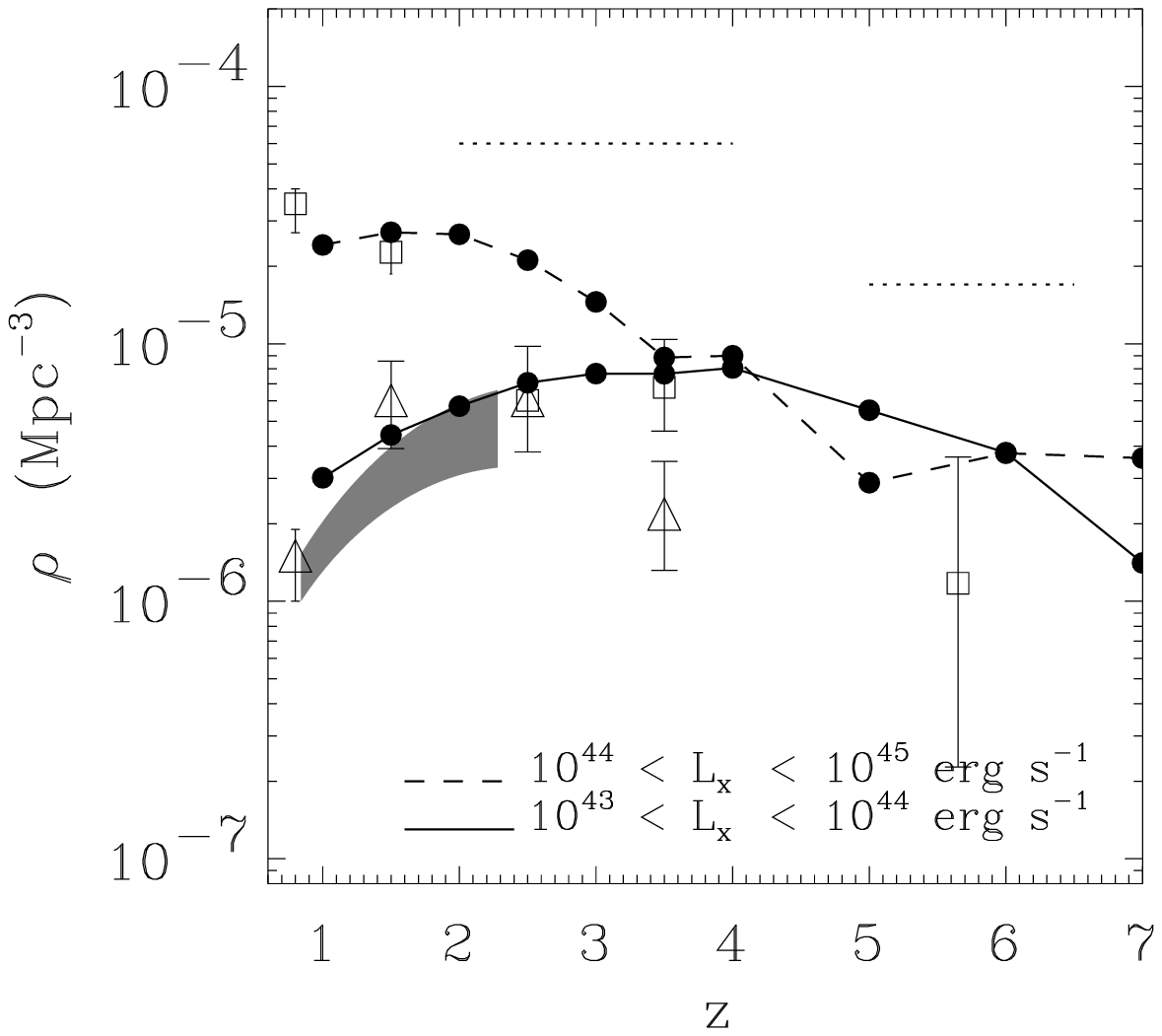,width=12.0truecm}}}
\caption{Top: The evolution of the space density of quasars with $M_{\rm
B} < -26$. Our model is shown by the solid circles joined by the thick
solid lines. The model is for $t_{Q} = 2 \times 10^{7} \yr$.  The
dotted line is a fit to the SSG survey in the redshift range $2.75 < z
< 4.75$. The open squares show a compilation from Pei (1995) derived
from the WHO and HS surveys.  The filled circles are the SDSS points
derived by Fan et al.~(2001a,b). The shaded area at low redshift is
the best fit model from the 2dF survey by Boyle et al. (2000). Bottom:
The evolution of space density of quasar in the $2-8 \keV$ X-ray
energy band. The data are from recent Chandra observations by Cowie et
al. (2003); Barger et al. (2003). The open squares represent the quasar
number density of sources with $10^{43} < L_{x} < 10^{44} \ergps$, and
the open triangle the number density of sources with $10^{44} < L_{x}
< 10^{45} \ergps$. The dotted bars show  upper limits. 
The solid and dashed lines show our model in these
luminosity ranges using with $t_{Q}=2\times 10^7 \yr$ and $t_{Q}= 10^7
\yr$ respectively. }
\end{figure} 

\subsubsection{X-ray band}
The bottom panel of Figure 8 shows the redshift evolution of the
comoving number density of quasar in the $2-8\keV$ band compared to
the recent Chandra observations (Cowie et al. 2003; Barger et
al. 2003). We plot our model with solid dots joined by a dashed line and
a solid line for two observed luminosity ranges $10^{43} < L_{x} <
10^{44} \ergps$ and $10^{43} < L_{x} < 10^{44} \ergps$ respectively.
We take $L_{x, 2-8 \keV} \sim 0.03 L$ where $L$ is the bolometric
luminosity given in Equation~\ref{eqn:Lacc}. This relation seems to be
roughly valid for all type of AGN (Elvis et al. 1994).  For the higher
luminosity range we take $t_{Q} = 2 \times 10^{7} \yr$ (as in \S 4.2.1),
whereas for the lower luminosity range the quasar timescale is a
factor of 2 smaller, $t_{Q} = 1 \times 10^{7} \yr$, as this seems to
describe better the overall evolution of these objects.
Overall, we find the model to agree pretty well with the current
constraints from the X-ray band if we allow a small,  mass dependent
 variation of the typical accretion timescale. This again 
emphasizes the strong dependence of our predictions on this parameter.

\subsubsection{Dust, obscured AGN, low-radiative efficiency}
Observations indicate that the comoving number density of quasars
peaks at $z \sim 3$ and drops abruptly by a factor of 20 from redshift
$z \sim 2$ to $z \sim 5$. Our predicted number density lies closer to
an extrapolation of the low redshift luminosity function (e.g., Boyle
et al. 2000), implying a perhaps less rapid evolution at high
redshifts. From the observational point of view, there is still some
uncertainty concerning the recent high-redshift results as
a result of a number of potentially very important systematic biases
(see e.g., Pei 1995; Fan et al. 2001a). Surveys at high redshifts
still suffer from incompleteness, uncertainties in the K-corrections
(when compared to low redshifts) and in particular from the
possibility that a large number of quasars are not detectable in
optical surveys as a result of dust extinction. 

The tendency to overestimate the number of bright quasars at $z
\approxlt 1$ in our model may also be less severe given the recent
{\em Chandra} results, which imply the presence of a substantial
population of optically obscured luminous AGNs at low redshifts (e.g.,
Rosati et al. 2002; Barger et al. 2001). This would help to reduce the
discrepancy. 
We also note that there is significant evidence from {\em Chandra}
observations of nearby galactic nuclei indicating that the gas supply
onto their central black holes far exceeds their luminosity outputs,
implying that the radiative efficiency $\eta$ of nearby supermassive
black holes may be a lot lower than the canonical value of $\eta =
0.1$ we have taken (e.g., the case of M87; Di Matteo et al. 2003; our
Galactic Center, Baganoff et al. 2002; Narayan 2002).

In summary, we have shown that the evolution of the gas supply via
star formation and feedback can, for the most part and without the
introduction of any additional parameters, reproduce the turnover in
the comoving number density of quasars at $z < 3$.  We obtain an
improved agreement with observations with respect to previous models
in which the evolution of the luminosity function is predicted within
the context of hierarchical build-up of the dark halos through merger
trees (e.g., Kauffmann \& Haehnelt 2000; Cattaneo 2000; Volonteri et
al. 2002; Hatziminaoglou et al. 2002). However other factors such as
obscuration or change in radiative efficiency or other feedback
processes (e.g.; due to radio jets) may still play some role in
producing the very sharp decrease in number counts at $z \approxlt 1$.

\begin{figure}[t]
\center{
\vbox{
\psfig{file=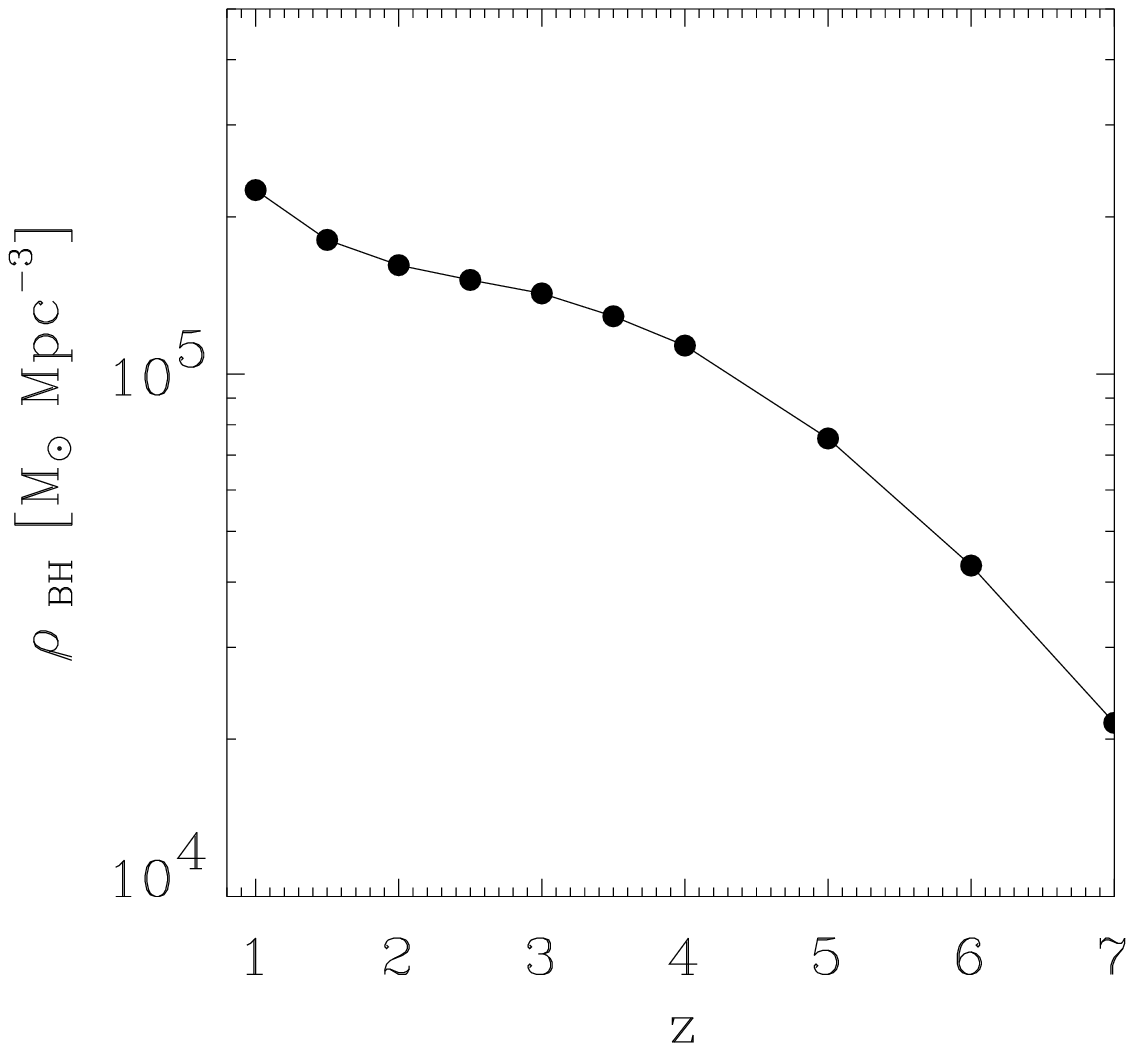,width=12.0truecm}
\psfig{file=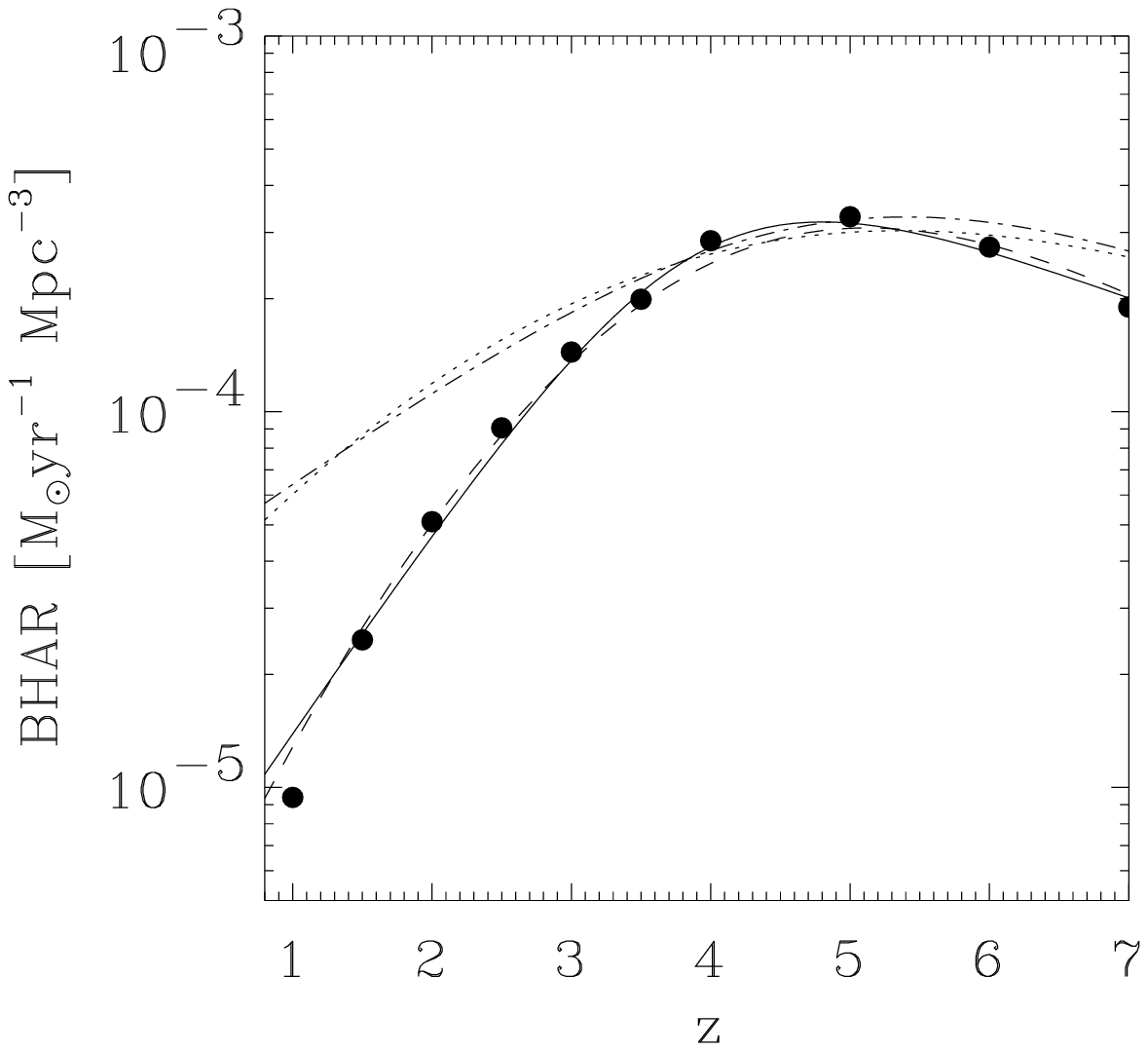,width=12.0truecm}}}
\caption{Top panel: The evolution of the black hole mass density in
our model. The majority of black hole mass is assembled by $z \sim
4$. Bottom panel: The evolution of the black hole accretion rate
(BHAR) density in our model for black hole activity described in
\S4. The solid circles mark the measurements from the simulations. The
solid line is an analytical fit to our results using the fitting
function of equation (8) and the dashed line using the function of
equation (9) where the numerator is $\propto \chi^4$ instead of
$\chi^2$. The dashed-dotted and dotted lines show the fits to the
cosmic star formation history in the simulation set analyzed by
Springel \& Hernquist (2003b) and Hernquist \& Springel (2003), but
scaled down by a factor of $500$. In our model, the black hole
accretion rate density has a peak at similar redshifts as the star
formation rate density, but it decreases faster at low redshifts.}
\end{figure}
 
\section{The black hole accretion history}

The evolution of the mass density of black holes predicted from our
model is shown in the top panel of Figure 9. Extrapolating to redshift
$z \sim 0$, we obtain $\rho_{\rm BH} \sim 2-4 \times 10^5 \Msun
\Mpc^{-3}$, which is within $20-30\%$ of the value obtained by Merritt
\& Ferrarese (2001), and is roughly consistent with the result of
$\rho_{\rm BH} = (2.5 \pm 0.4)\times 10^5 \Msun \Mpc^{-3}$ derived by
Yu \& Tremaine (2002) using the velocity dispersions of galaxies in
the SDSS (also similar to the value of Salucci et al. 1999).
Values inferred from the X-ray background lie typically somewhat
higher, in the range of $\rho_{\rm BH} = 6 - 16.8
\times 10^5 \Msun \Mpc^{-3}$ (Fabian \& Iwasawa 1999; Elvis, Risaliti
\& Zamorani 2002). Although the most recent X-ray constraint from Chandra 
observations give $\rho_{\rm BH} \sim 2-4 \Msun \Mpc^{-3}$ (Cowie et
al. 2003). Note that Figure 9 clearly shows that the majority of the
black hole mass is already built up by $z \sim 4$.

In the bottom panel of Figure 9, we plot the evolution of the comoving
black hole mass accretion rate density according to our model. As
suggested by the cumulative black hole mass function in the top panel,
the peak of the black hole accretion rate per unit volume occurs at
redshifts around 4-5.

It is also interesting to compare the black hole accretion rate
density (BHAR) directly with the cosmic star formation rate density
(SFR), as measured by Springel \& Hernquist (2003b) for the simulation
set we analyze here.  The dashed line in the bottom panel of Figure 9
shows an empirical double-exponential fit used by Springel
\& Hernquist (2003b) to approximately describe the simulation results
for star formation, here rescaled by a factor $2 \times 10^{-3}$.

Comparing this fit with our results for black hole growth, it is
evident that the shapes of the BHAR and SFR histories are quite
similar in general, particularly at high redshifts, but that the BHAR
history more steeply declines at low redshifts.

We may hence also try to use the double exponential function employed
by Springel \& Hernquist (2003b) to fit the BHAR. Using the ansatz
\beq
\dot{\rho} (z) = \epsilon \frac {b
\exp[a(z-z_{m})]}{b-a+a\exp[b(z-z_{m})]}, \label{eqn_fit1}
\eeq
we obtain a good fit for our measurements of the BHAR with parameter
values $a=5/4$, $b=3/2$, $z_m = 4.8$ and $\epsilon=\epsilon_{\rm BH}=
3 \times 10^{-4} \Msunpyr\Mpc^{-3}$.  This fit is shown as the solid
line in the bottom panel of Figure 9.

Compared to the SFR history, the peak of the BHAR appears to occur at
slightly lower redshift, but the high-z evolution is nevertheless
broadly similar, suggesting that the evolution of both rates is
largely driven by the rapid gravitational growth of the halo mass
function in this regime.  Note however that even at its peak, the
black hole accretion rate density is roughly a factor 500 lower than
the star formation density, implying that the rate at which baryons
are locked up in stars always far exceeds that in black holes.

It is also interesting to compare our results and the differences
between the SFR and BHAR in terms of the analytic model given by
Hernquist \& Springel (2003) for the cosmic star formation history.
Using detailed analytical arguments based on the dependence of the SFR
on cosmological growth of structure and the physics of star formation,
they {\em derived} a fitting function for the SFR history given by
\beq 
\dot{\rho}_{\ast} (z) = \dot{\rho}_{\ast}(0) \frac{\chi^2}{1+\alpha (\chi -1)^3
\exp(\beta \chi^{7/4})},
\label{eqn_fit2}
\eeq
where $\chi \equiv [H(z)/H_{0}]^{2/3}$, and where $\alpha$, $\beta$
and $\dot{\rho}(0)$ are parameters. In fact, Hernquist \& Springel
(2003) showed that this form (also drawn as the dotted line in Fig. 9)
provides a more accurate fit to the simulation results for the star
formation rate than equation (\ref{eqn_fit1}).

In particular, Hernquist \& Springel (2003) demonstrated that the low
redshift dependence of the star formation rate, $\dot{\rho}_{\ast} (z)
\propto H(z)^{4/3}$, is primarily caused by the declining efficiency
of gas cooling, which itself is related to the expansion rate of the
universe, as measured by the evolution of the Hubble constant, $H(z)$.
At high redshifts on the other hand, the evolution is driven by the
gravitational growth of structure. It is therefore not too surprising
that the BHAR closely matches the SFR at high redshifts, but shows a
different behavior at low redshifts, since we assumed that black hole
accretion is not linked in the same way as star formation to gas
cooling, but instead to stellar feedback processes, most notably to
galactic outflows. In the bottom panel of Figure 9 we also show a fit
to the BHAR using a function similar to equation (\ref{eqn_fit2}) where,
in order to fit the steeper decline at intermediate to low redshifts we
have assigned the numerator to be $\propto \chi^4$ instead of
$\propto \chi^2$ (and $\alpha = 0.013$, $\beta=0.12$ and $\rho_{\rm BH}
(0)= 2.2 \times 10^{-4} \dot{\rho}_{\ast}(0)$). At intermediate to low
redshift the BHAR therefore scales as $\dot{\rho}_{BH} \propto
\dot{\rho}_{\ast} (z) \times H(z)^{4/3} \propto H(z)^{8/3}$;
i.e. roughly as the square of the cooling rate. This new scaling may
arise as a result of the fact that gas is depleted by the joint
effects of star formation and feedback. The BHAR then may scale in
proportion to both the star formation (which scales as $H(z)^{4/3}$ in
this regime) and the feedback processes (which themselves scale as the
SFR), hence as the square of the cooling rate. In future work, we will
investigate in more detail possible physical processes underlying this
scaling by means of analytical arguments.

\section{Conclusions}
We have explored the hypothesis that black hole growth and activity at
the centers of galaxies is primarily regulated by the evolution of the
total baryonic gas mass content of galaxies. We have used $\Lambda$CDM
cosmological hydrodynamic simulations, which include a converged
prescription for star formation, to follow the evolution of the
baryonic mass in galactic potential wells. Using the numerical models,
we have searched for correlations in the large-scale properties of
galaxies that match those of the observed black holes. Note that
with such an approach, resolving the properties of the actual small-scale
accretion flows around black holes is not important.  

We have shown that the steep power law form of the observed $M_{\rm
BH}- \sigma$ correlation can be understood over an interesting regime
of circular velocities and redshifts if there is a simple linear
relation between the total gas mass (subject to star formation and
associated feedback) in galaxies and their black hole mass.  We find
that the observed black hole mass density is consistent with $M_{\rm
BH} \sim 0.004 h^{-1} M_{\rm gas}$; i.e. the central black holes
contain $\approxlt 1\%$ of baryons in galaxies. The total amount of
gas in galaxies, and hence their black hole mass, may saturate simply
in response to star formation, and in particular to supernova feedback
and galactic winds. By considering these as the only physical
processes regulating the growth of black holes, we predict that the
steep dependence of black hole mass on velocity dispersion is tight
but not set in primordial structures, but only fully established at
low redshifts, $z \approxlt 2$. Once the relationship is established,
we find that the black hole mass is related to the dark matter mass by
$M_{\rm BH}/{10^8 \Msun} \sim 0.7 (M_{\rm DM} / 10^{12} \Msun)^{4/3}$.

We assume that all galaxies undergo a quasar phase with a typical
lifetime, $t_{Q} \sim$ a few $\times 10^7$ yr (of the order of the
Salpeter time) and show that the evolution of the quasar gas supply
(taken as the $\sim 1\%$ fraction of the total gas) in spheroids is
sufficient to explain, for the most part, the decrease of the bright
quasar population at redshift $z <3$.  However, at redshift $z\sim 1$
the predicted shape of the luminosity function differs from
observations implying that mass dependent variations in the typical
accretion timescale or radiative efficiency are likely to be
important.  An additional possibility is that there is significant
intrinsic obscuration (as suggested by recent {\em Chandra}
constraints on the hard XRB; e.g., Rosati et al. 2002) which may also
be mass dependent.

The quasar number density at high redshift in our model does decline,
 but under the simplest assumption of a redshift independent $t_{q}$,
 it can still exceed the SDSS space density of bright quasars at $4
 \approxlt z \approxlt 6$ (e.g., Fan et al. 2001a,b) by a factor $\sim
 2$ or more.  However the observed $B$-band number densities at $z
 \approxgt 4$ may be significantly biased low by dust along the line
 of sight (Pei 1995; Fan et al. 2001a), as is suggested by
 observations of high redshift quasars which are found to reside in
 dusty environments (e.g., Omont et al. 1996; Carilli et al.~2000,
 2001).  These biases in in B-band selection can be circumvented by a
 comparison of our predictions with the newly constrained comoving
 number density of X-ray selected quasars (Cowie et al. 2003). We have
 done this and find that the model yields the observed evolution of
 number density of X-ray faint quasars over the whole redshift range
 $1 < z< 6$ although the statistical significance of these
 observations is still not very high.

We further note that any freedom we have to change the luminosity
function of quasars in our model comes down to our ability to select
the accretion timescale $t_{Q}$, which is our only free parameter in
the model.  This is a fundamental parameter which governs black hole
and quasar evolution, and observational constraints on its value will
make our model predictions more firm and comparisons with data even
more interesting. In the next few years constraints on quasar
lifetimes are likely to become available from study of the clustering
and the proximity effects of quasars mostly from the SDSS.

It has been shown that models based on the Press-Schechter formalism,
where the quasar luminosity function is related to the dark matter
halo mass, and quasar emission is assumed to be triggered by galaxy
mergers, can reproduce the statistical properties of the observed SDSS
quasar luminosity functions if appropriate assumptions for the quasar
duty cycles are made (e.g., Wyithe \& Loeb 2002; Volonteri et al. 2002;
Hatziminaoglou et al. 2002). However, in these models, the decrease in
merging rates is not sufficient to reproduce the luminosity functions
at low redshifts and additional modeling of the evolution of the gas
fraction is then required (Kauffmann \& Haehnelt 2000). This
highlights the importance of properly including the physical processes
that govern the evolution of baryons trapped in halos as well as the
growth of galaxies and their interactions. In this work we did not
assume that quasar emission is triggered by galaxy mergers, but
simply assumed that all galaxies that are
star-forming\symbolfootnote[1]{Note this choice does not affect our
results. It simply excludes small-mass objects where star formation is
likely not to be resolved.} have an equal probability of hosting a
quasar. Because the quasar lifetime $t_{Q}$ is short, there can be
many generations of quasars in our model, but the number density of
quasar hosts is significantly larger at high redshifts, where $t_{Q}$
becomes a more significant fraction of the age of the universe than
at low redshifts.

We have used our model for the relation between gas mass and black
hole mass to predict the evolution of the black hole mass density in
galaxy centers. We have shown that if black hole mass is assembled by
gas accretion, our predicted value for the black hole mass density
$\rho_{\rm BH} \sim 2-4 \times 10^5 \Msun\Mpc^{-3}$ (extrapolated to
$z=0$) is consistent with observational estimates for the local value
(Yu \& Tremaine 2002). We find that the large majority of black hole
mass is assembled up to and preceding the peak of the bright quasar
phase at $z \sim 3$ and that almost no further growth takes place at
lower redshifts.

We have also derived the evolution of the black hole accretion rate
density (BHAR) and showed that the majority of black hole accretion
should occur in the high density environments at $z \sim 4 -5$, in
rough correspondence with the peak of the star formation rate (SFR)
history in the simulations. This is consistent with the derived black
hole mass density evolution and implies that the peak of the optically
bright quasar phase occurs only when the largest black holes are
already assembled. At very high redshift, the BHAR and the SFR evolve
similarly, both primarily driven by the rapid gravitational growth of
structure, while at low redshift, the BHAR declines much more rapidly.
Although the peak of the BHAR occurs close to that of the star
formation rate, its normalization is a factor of a few hundred lower
than that of the cosmic SFR.  It hence seems unlikely that black hole
accretion plays a crucial role for the gas dynamics in galaxies, and
it may even be relatively unimportant for the general process of
galaxy formation, unless there is strong energetic feedback by
active QSOs that affects galaxies in a significant way.
In future work we will investigate the predictions of this
model for very high redshifts and in particular its implication
for the nature of the ionizing background.

\acknowledgements

This work was supported in part by NSF grants ACI 96-19019, AST
98-02568, AST 99-00877, and AST 00-71019.  The simulations were
performed at the Center for Parallel Astrophysical Computing at the
Harvard-Smithsonian Center for Astrophysics.



\end{document}